\documentclass[twocolumn,aip,reprint,numerical]{revtex4-2}
\usepackage{amsfonts}
\usepackage{graphicx}
\usepackage{epstopdf}
\usepackage{epsfig}
\usepackage{amsmath}
\usepackage[normalem]{ulem}
\usepackage{multirow}
\usepackage{makecell}
\usepackage{array}
\usepackage{booktabs}
\usepackage{mathtools}
\usepackage{bm}
\usepackage{color}
\usepackage{soul,xcolor}
\usepackage{array}
\usepackage{booktabs}

\usepackage{tabularx}
\newcolumntype{Y}{>{\centering\arraybackslash}X}

\usepackage{blindtext}
\usepackage{comment}
\definecolor{lightblue}{rgb}{0.3, 0.3, 0.90}
\definecolor{darkblue}{rgb}{0.0, 0.2, 0.5}
\definecolor{lightred}{rgb}{1.0, 0.6, 0.6}
\definecolor{darkred}{rgb}{0.8, 0.0, 0.0}

\begin{document}
\title{Entanglement engineering of optomechanical systems by reinforcement learning}

\author{Li-Li Ye }
\affiliation{School of Electrical, Computer and Energy Engineering, Arizona State University, Tempe, Arizona 85287, USA}
\date{\today}

\author{Christian Arenz}
\affiliation{School of Electrical, Computer and Energy Engineering, Arizona State University, Tempe, Arizona 85287, USA}

\author{Joseph M. Lukens}
\affiliation{Research Technology Office and Quantum Collaborative, Arizona State University, Tempe, Arizona 85287, USA}
\affiliation{Quantum Information Science Section, Oak Ridge National Laboratory, Oak Ridge, Tennessee 37831, USA}

\author{Ying-Cheng Lai} \email{Ying-Cheng.Lai@asu.edu}
\affiliation{School of Electrical, Computer and Energy Engineering, Arizona State University, Tempe, Arizona 85287, USA}
\affiliation{Department of Physics, Arizona State University, Tempe, Arizona 85287, USA}

\begin{abstract}

Entanglement is fundamental to quantum information science and technology, yet controlling and manipulating entanglement --- so-called entanglement engineering --- for arbitrary quantum systems remains a formidable challenge. There are two difficulties: the fragility of quantum entanglement and its experimental characterization. We develop a model-free deep reinforcement-learning (RL) approach to entanglement engineering, in which feedback control together with weak continuous measurement and partial state observation is exploited to generate and maintain desired entanglement. We employ quantum optomechanical systems with linear or nonlinear photon-phonon interactions to demonstrate the workings of our machine-learning-based entanglement engineering protocol. In particular, the RL agent sequentially interacts with one or multiple parallel quantum optomechanical environments, collects trajectories, and updates the policy to maximize the accumulated reward to create and stabilize quantum entanglement over an arbitrary amount of time. The machine-learning-based model-free control principle is applicable to the entanglement engineering of experimental quantum systems in general. 

\end{abstract}

\maketitle

\section{Introduction} 

Entanglement~\cite{bennett:2000,Braunstein:2005,Huang:2021,huang:2022,lee:2023} is 
fundamental to all fields in quantum information science such as quantum 
sensing~\cite{degen:2017}, quantum computation~\cite{horodecki:2009}, and quantum
networks~\cite{kimble:2008,wehner:2018,bouwmeester:1997,ren:2017,chen:2021,pirandola:2017}. 
However, the inherent fragility of quantum entanglement and coherence~\cite{de:2021} 
poses significant challenges for experimental applications. For example, in quantum 
computing, the application of quantum gates to quantum states needs to last for a 
finite amount of time~\cite{noiri:2022,magann:2021,banchi:2016,romero:2012,van:2012,chow:2012}, 
making it critical to maintain the entanglement after its creation. Moreover, the 
transition from noisy intermediate-scale systems~\cite{bharti:2022} to large-scale, 
fault-tolerant systems~\cite{magann:2021} requires sophisticated entanglement 
engineering strategies to establish and maintain entanglement through optimal control
protocols in the presence of noise and decoherence.

At the present, a major limitation/challenge in entanglement engineering is 
the experimental observation design. Existing machine-learning based works
use the full fidelity, i.e., the overlap between the current and target quantum 
states, as the observation metric. Applications range from the generation 
of two~\cite{mackeprang:2020} and multi-qubit entangled 
states~\cite{giordano:2022,zhang:2024} to specific many-body 
states~\cite{metz:2023,bukov:2018,guo:2021} and single-particle quantum state 
engineering via deep reinforcement learning (RL)~\cite{sivak:2022,porotti:2022}. 
However, full fidelity observation is not universally applicable in experiments. 
Moreover, obtaining the relationship between the entanglement and experimental 
observables is difficult. So far there have been no systematical methods to 
extract quantitative entanglement from experimental observation for arbitrary 
quantum systems~\cite{jayachandran:2023,ho:2018,gut:2020}, in spite of some 
initial exploration for specific systems. For example, an entanglement criterion 
for non-Gaussian states in coupled harmonic oscillators was 
developed~\cite{jayachandran:2023}. Under the strong laser approximation, a 
Bell inequality was tested with photon counting~\cite{ho:2018}, and stationary 
entanglement for Gaussian states was inferred from the continuous measurement of 
light only~\cite{gut:2020}. Recently, conditional state tomography~\cite{borah:2023} offers a method to reconstruct the density matrix from the weak measurement current. However, it requires the use of a specific Hamiltonian form, and the finite time needed for the conditional density matrix to converge to the actual density matrix. This highlights that quantum state tomography in a model-free manner for measurement-based feedback control remains a significant challenge.

In this paper, using quantum optomechanical systems with linear or nonlinear 
photon-phonon interactions as a paradigm, we develop a deep RL approach to 
entanglement engineering. For quantum control of optomechanical systems, most 
existing theoretical studies focused on Gaussian states or the linear 
interaction regime~\cite{hofer:2011,cai:2023,clarke:2020,kiesewetter:2014,wang:2014,hofer:2015,lin:2014,farace:2012,mari:2012,mari:2009,stefanatos:2017,hofer:2015,guo:2022,harwood:2021}, 
with the primary goal of generating entanglement as quickly as possible 
(entanglement enhancement)~\cite{stefanatos:2017,hofer:2015,guo:2022,harwood:2021}. 
Previous control methods are mostly model-based: prior information about the system 
model is needed, such as the pulse 
method~\cite{hofer:2011,cai:2023,clarke:2020,kiesewetter:2014,wang:2014}, 
time-continuous laser-driven approaches~\cite{hofer:2015,lin:2014}, periodic 
modulations~\cite{farace:2012,mari:2012,mari:2009}, optimal pulse 
protocols~\cite{stefanatos:2017}, linear quadratic-Gaussian (LQG) 
methods~\cite{hofer:2015}, and coherent feedback methods using auxiliary optical 
components~\cite{guo:2022,harwood:2021}. We note that there were two previous 
works~\cite{wang:2020,Borah:2021} on model-free RL for controlling and stabilizing 
a quantum system with an inverted harmonic potential and a double-well nonlinear 
potential, respectively, to a target state using weak-current measurements (WCMs) 
and partial state observation. However, these two works did not address 
entanglement control, while our work is developing a model-free deep-RL method to 
realize non-Gaussian entanglement engineering using only photon number counting from 
WCMs. (Backgrounds about WCM, deep RL, and quantum control are presented in 
Appendix~\ref{sec:appendix_background_work}.) To our knowledge, prior to our work, 
model-free deep RL feedback control to create and stabilize the entanglement with 
WCM observations had not been available. 

The particular aspects of our work that go beyond the existing works are briefly
described, as follows. In our work, in the linear (nonlinear) interaction regime, 
the observation is the WCM photocurrent (the expectation value of the photon number). 
We note a previous work~\cite{porotti:2022} that employed a proximal policy 
optimization (PPO)~\cite{schulman:2017} RL agent, to generate different Fock 
states and the superposition of a single cavity mode based on observing the 
density matrix and a fidelity-based reward function. In contrast, the observable 
in our work is the photocurrent that is more experimentally 
accessible~\cite{essig:2021}. For quantum measurement, we use WCM in real-time 
feedback control, taking into consideration the resulting quantum stochastic 
process~\cite{Smith:2004,jacobs:2006,Borah:2021}, and identify a numerical 
relationship between the entanglement and photocurrent. In both the linear and
nonlinear regimes, we focus on non-Gaussian state control because, according
to the nonlinear quantum master equation resulting from WCM, the time evolving 
quantum states are intrinsically non-Gaussian. Our deep RL control scheme is 
model-free~\cite{ramirez:2022}, where policies or value functions are directly 
learned from the interactions with the quantum environment without any explicit 
model of this environment. This should be contrasted to the model-based deep RL 
methods~\cite{kaiser:2019}, where a pre-built model of the environment for policy 
decision-making is needed. We demonstrate that, under the actions of the 
well-trained PPO or recurrent PPO RL agent, entanglement between the quantum 
optical and mechanical modes can be created and maintained about the target 
entanglement. 

Our main results are as follows. First, under the strong laser 
approximation, the interaction resulting from the radiation pressure between 
the cavity and the mechanical oscillator modes can be linearized and described 
by the beam-splitter Hamiltonian. During the training phase, the PPO agent 
interacts with parallel quantum environments and collects the subsequent data 
by episodic learning, with the observation being the WCM photocurrent. The 
deep-RL method can extract useful information from the measurement photocurrent, 
which is encoded in the Wiener process, and achieve the target entanglement 
engineering in a model-free manner for the quantum system that is dissipative 
due to coupling to the vacuum bath and is driven by a laser. In the testing 
phase, with the agent interacting and observing a single quantum environment,
we demonstrate that the entanglement-engineering performance of our 
deep-RL method with WCM observation greatly exceeds that of both state-based 
Bayesian methods~\cite{Stockton:2004, Borah:2021} and random control. Second,
when the driving laser field is not strong, the quantum optomechanical 
interaction is nonlinear~\cite{liu:2013,qian:2012}. In this case, we articulate 
two training phases for nonlinear entanglement engineering. The first phase is 
utilized to infer the entanglement by the model-free deep RL, dubbed as the 
target-generating phase, where the observation of the PPO agent [with 
multilayer perceptions (MLPs)] is the logarithmic negativity and the reward 
function is constructed to limit the high-level excitation and facilitate 
entanglement learning. (Direct experimental measurement of the logarithmic 
negativity is currently not available.) The time series of the expected 
photon number in the regime from converged training episodes is selected as 
the target for the next phase. The second phase is then the target-utilization 
phase, where the recurrent PPO (with long short-term memory 
(LSTM)~\cite{hochreiter:1997} added after MLPs) observes the expected photon 
number and obtains the reward only based on the target expected photon number 
obtained from the last phase. In this framework, the recurrent PPO controls 
the quantum state in the low-energy regime with the desired entanglement 
created and stabilized.

\begin{figure*} [ht!]
\centering
\includegraphics[width=0.9\linewidth]{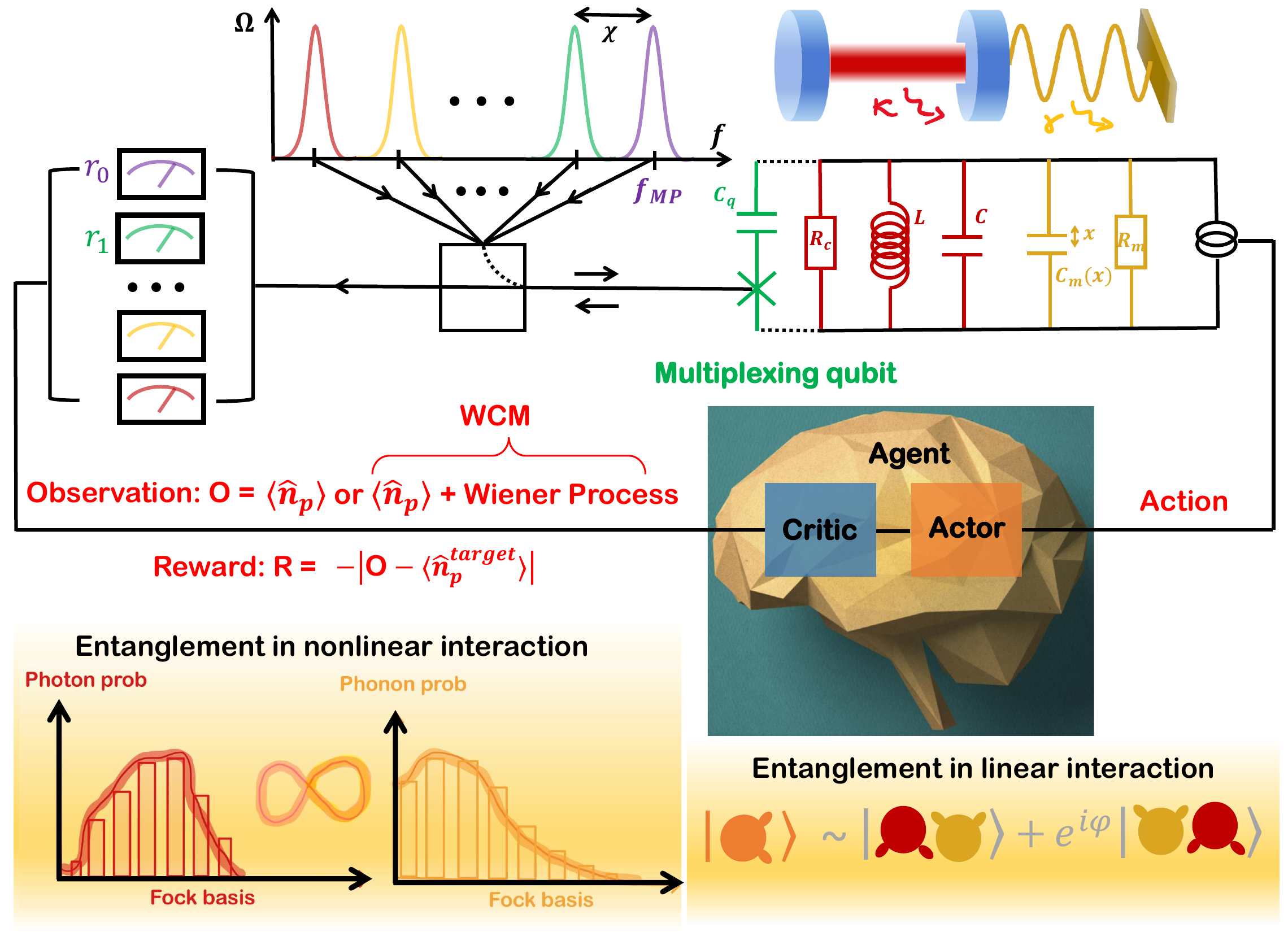}
\caption{Experimental proposal of measurement-based feedback control of deep RL 
to create and stabilize entanglement in an open quantum optomechanical system 
dissipatively coupled to the vacuum bath. Quantum optomechanics was experimentally 
realized in a microwave electromechanical 
system~\cite{dobrindt:2008,palomaki:2013,barzanjeh:2022}, where the multiplexing 
qubit was used to weakly couple to the microwave resonator for extracting the 
photon number statistics through weak measurements~\cite{essig:2021}. The RL agent 
acts in one or multiple parallel quantum optomechanical environments according to 
the parameterized policy and collects data in one episode consisting of $T$ time 
steps: observations $O_t$, reward $R_t$, and actions, after which the quantum 
optomechanical environment is reset. After one or several episodes, the policy is 
updated using minibatch data to maximize the accumulated reward. The aim is to 
achieve the desired entanglement $E_N\sim \log 2\sim 0.7$ (in the natural 
logarithmic base) between the cavity-optical and mechanical modes. Entanglement 
engineering of this type can be achieved in both the linear and nonlinear 
interaction regimes. In the linear case, the task is similar to that of achieving 
an entangled Bell state of the beam-splitter Hamiltonian or ``swap'' Hamiltonian. 
In the nonlinear regime, the entangled states from entanglement engineering can
be complicated. Illustrated are the resulting photon and phonon number distributions 
of the entangled states.}
\label{fig:scheme}
\end{figure*}

\section{Results}

\subsection{Experimental proposal for entanglement engineering}

Our goal is achieving entanglement engineering between the optical cavity and 
mechanical oscillator modes using deep RL. Based on the current experimental 
progress, we articulate an experimental proposal to achieve this goal, as shown
in Fig.~\ref{fig:scheme}. Consider a Fabry-Perot cavity that consists of a 
single-mode cavity and a movable end mirror. The optical cavity has the 
frequency $\omega_c$ and the optical field exerts a radiation pressure on the 
mirror. The cavity mode is driven externally by a coherent laser field with 
frequency $\omega_L$. The mirror's quantized center-of-mass motion is described 
by a harmonic oscillator of frequency $\omega_m$. In the rotating frame of the 
laser, the Hamiltonian describing the coupling between the optical cavity and 
mechanical oscillator modes is given by~\cite{liu:2013,qian:2012}
\begin{align} \nonumber
\tilde{H}_{nl} = -\hbar\Delta \hat{a}^{\dagger}\hat{a} + \hbar\omega_{m}\hat{b}^{\dagger}\hat{b}+\hbar g_0(\hat{b}^{\dagger}+\hat{b})\hat{a}^{\dagger}\hat{a}+\hbar\alpha_{L}(\hat{a}^{\dagger}+\hat{a}),
\end{align}
where $\hat{a}$ and $\hat{b}$ are the annihilation operators of the cavity and 
mechanical mode, respectively, $\hat{a}^{\dagger}$ and $\hat{b}^{\dagger}$ are 
the corresponding creation operators. The frequency detuning of the cavity is 
$\Delta \equiv \omega_L-\omega_c$. The nonlinear coupling $g_0$ arises from the 
radiation pressure force between the light and the movable mirror (details given 
in Appendix~\ref{sec:appendix_Hamiltonian}), and $\alpha_L$ is the real amplitude of the driven 
electromagnetic field. We set $g_0>\kappa$ so that the single-photon optomechanical 
coupling rate $g_0$ exceeds the coupling strength $\kappa$ between the cavity and 
the vacuum bath. This condition guarantees observable nonlinear quantum 
effects~\cite{Aspelmeyer:2014}. Under the strong laser approximation: 
$|\bar{\alpha}_c|\gg1$, where $|\bar{\alpha}_c|$ is the amplitude of the light 
field inside the cavity induced by the strong laser, we have
$\hat{a}\approx \bar{\alpha}_c+\delta\hat{a}$ with $\delta\hat{a}$ denoting the 
excitation or the shifted oscillator on top of the large coherent state with 
the amplitude $\bar{\alpha}_c$. After the displacement transformation $e^{\bar{\alpha}_{c}\hat{a}^{\dagger}-\bar{\alpha}_{c}^{*}\hat{a}}\hat{a}e^{-\bar{\alpha}_{c}\hat{a}^{\dagger}+\bar{\alpha}_{c}^{*}\hat{a}}=\hat{a}-\bar{\alpha}_{c}\equiv\delta\hat{a}$, the resulting linearized beam-splitter or 
``swap'' Hamiltonian ~\cite{lemonde:2013,liu:2013} is
\begin{align} \nonumber
\tilde{H}_{bs}\approx\hbar\omega_{m}\delta\hat{a}^{\dagger}\delta\hat{a}+\hbar\omega_{m}\hat{b}^{\dagger}\hat{b}+\hbar G(\delta\hat{a}^{\dagger}\hat{b}+\hat{b}^{\dagger}\delta\hat{a}),
\end{align}
which is obtained in the red-detuned regime $\Delta=-\omega_{m}$, where the 
coefficient $G\equiv g_0\bar{\alpha}_c$ can be tuned by the amplitude of the 
incoming laser (a time-dependent modulation)~\cite{florian:2014}. The interaction 
term describes the state transfer between photons and phonons in the strong coupling 
regime for $G>\kappa$, with $\kappa$ ($\gamma$) being the decay rate of the cavity 
(mechanical) mode to the vacuum bath at zero temperature. 

Our control strategy was developed based on considering the current experimental 
capability. Previous works on the microwave regime of the optomechanical 
systems~\cite{dobrindt:2008,palomaki:2013,barzanjeh:2022} suggested the feasibility
of the experimental implementation of our RL control scheme. In particular, a 
one-to-one correspondence between the Fabry-Perot cavity and the microwave 
electromechanical system was 
demonstrated~\cite{song:2014,palomaki:2013,dobrindt:2008}. As shown in 
Fig.~\ref{fig:scheme}, the microwave resonator of an LC circuit is equivalent to 
the Fabry-Perot optical cavity mode with the movable capacity~\cite{song:2014} 
$C_m(x)$ corresponding to the flexible mirror in the optical cavity. The resistors 
$R_c$ and $R_m$ can be related to the decay rate $\kappa,\gamma$ to the vacuum 
bath~\cite{song:2014}. Based on the experimental results, we can compare the 
typical parameter configurations between the optomechanical and electromechanical 
systems. The decay rate of the optical cavity mode is $\kappa=0.01\,\omega_m$ in 
the linear regime and $\kappa=0.1\,\omega_m$ in the nonlinear regime, with the 
better quality of the mechanical oscillator mode $\gamma = 0.01\,\kappa$. 
Consequently, we have $\gamma\approx10^{-3}\,\omega_m\sim 10^{-4}\,\omega_m$. 
The typical experimental decay rate of the microwave resonator 
is~\cite{palomaki:2013,dobrindt:2008,shin:2022,seis:2022,liu:2024,Aspelmeyer:2014}
$\kappa\approx 0.01\,\omega_m\sim0.1\,\omega_m$ with $\gamma\approx 10^{-3}\,\omega_m\sim 10^{-9}\,\omega_m$. In our work, the nonlinear coupling is set to be
$g_0=0.2\,\omega_m$, whereas the typical coupling in the strong coupling regime 
in a previous work~\cite{dobrindt:2008} was about $g_0=0.1\,\omega_m$. The strength 
of the laser in our work is $G\in[-5,5]\,\omega_m$ for the linear system in the 
red-detuned regime $\Delta = -\omega_m$ and $\Delta,\alpha_L\in[-5,5]\,\omega_m$
for the nonlinear system. In the microwave version, this range can be adjusted 
by the pump's strength~\cite{palomaki:2013,dobrindt:2008,shin:2022,seis:2022,liu:2024,Aspelmeyer:2014}.

In the microwave regime, it was demonstrated that the photon-number statistics of 
a microwave cavity mode can be detected using multiplexed photon number 
measurements~\cite{essig:2021,johnson:2010,gleyzes:2007}. By this method, the 
multiplexing qubit encodes multiple bits about the photon number distribution of 
a microwave resonator through dispersive interaction. A frequency comb drive,
distributed at $f_{\textnormal{MP}}-k\chi$, reads out all the information about 
the photon number distribution at once~\cite{essig:2021}, where $k$ denotes the 
number of photons and $\chi$ represents the dispersive qubit-resonator coupling, 
as shown in Fig.~\ref{fig:scheme}. The reduction in the reflection amplitude, 
$1-r_k$ with $k=0,1,...$, of the frequency comb, is proportional to the 
photon-number distribution of the microwave cavity mode over the Fock bases, as 
detected by the weak measurement~\cite{porotti:2022,essig:2021}. In our circuit 
design of experimental proposal, we add a capacitor $C_q$ to realize the weak 
coupling to the original electromechanical system. The coupling capacitance is 
small enough to be neglected in the total Hamiltonian, but it still allows the 
multiplexing qubit, denoted by the green cross in Fig.~\ref{fig:scheme}, to 
encode the photon number distribution of the microwave resonator through dispersive 
interaction.  

Under weak measurement~\cite{porotti:2022,essig:2021}, the sequence of the reduced 
reflection amplitude $1-r_k$ is collected by the PPO agent, which is proportional 
to the occupied photon number probability. Consequently, the expected photon number 
is calculated as 
\begin{align} \nonumber
\langle\hat{n}_p\rangle=\sum_{n} n\langle\sqrt{\eta}\hat{P}_n\rangle/\sqrt{\eta}=\sum_{n} n\langle\hat{P}_n\rangle 
\end{align}
and the WCM photocurrent is 
\begin{align}
\sqrt{\eta}\,\mathcal{I}(t)=\sum_{n}n\left[\langle\sqrt{\eta}\hat{P}_n\rangle+\frac{dW_n(t)}{\sqrt{4\eta}dt}\right]
\end{align}
with the measurement rate $\eta$, where $\hat{P}_n=|n\rangle\langle n|$ is the 
measurement projector on the Fock state $|n\rangle$, and $dW(t)$ is the Wiener 
increment with zero mean and variance $dt=0.01\,\omega_{m}^{-1}$ (the time step 
size in our calculations). In the linear quantum optomechanical regime, the Fock 
space for each mode is limited to $n=0,1$. The action is the amplitude modulation 
of the laser, which is in the range $G\in[-5,5]\,\omega_m$. In the nonlinear 
regime, the Fock dimension is $n=0,1,\ldots,9$. The time-dependent control signal 
consists of the detuning $\Delta$ and the amplitude $\alpha_L$ of the driven laser 
within the fixed range $\Delta,\alpha_L\in[-5,5]\,\omega_m$. 

The open dissipative quantum optomechanics under the WCM obey the stochastic 
master equation (SME) (see Sec.~\ref{sec:methods} and 
Appendix~\ref{sec:appendix_sme}). The number $n_{\textnormal{traj}}$ of 
trajectories simulated from SME can be selected according to the following 
considerations. If the observable is some expected physical quantity, using one 
trajectory is sufficient to extract the information about the quantum state: 
$n_{\textnormal{traj}}=1$. Experimentally, WCMs are performed, encoding the 
Wiener process in the observation and resulting in a large variance from the 
expectation value. To reduce the variance, more quantum trajectories should be 
used. To make computations feasible, we use five trajectories: 
$n_{\textnormal{traj}}=5$.

In the online training phase, for each episode with time steps, e.g., $T=500$, 
the PPO agent - the combination of the actor and critic network, collects the 
sequence of the observations $O(t)=\langle \hat{n}_p\rangle(t)$ or 
$\mathcal{I}(t)$, the reward value 
$R(t)=-|O(t)-\langle \hat{n}^{\textnormal{target}}_p\rangle(t)|$, and the resulting 
actions generated by its policy. After one or several episodes, the policy of the 
PPO agent is updated using minibatch data to maximize the accumulated reward. The 
RL agent is designed to interact with a single or multiple parallel quantum 
environments to make the time evolving observation $O(t)$ align with the target 
one $\langle \hat{n}^{\textnormal{target}}_p\rangle(t)$. In the online testing 
phase, the policy of the well-trained agent will not update and only interact with 
a single quantum environment to give the optimal control protocol to the 
corresponding observation. To realize entanglement engineering, i.e., achieving 
the desired entanglement between the cavity-optical and mechanical modes, finding 
the relation between the experimental observables and entanglement quantities is an 
unavoidable challenge. In our work, the model-free PPO agent finds the numerical 
relationship between them and realizes the entanglement engineering in both the 
linear and nonlinear regimes of quantum optomechanics, as shown in 
Fig.~\ref{fig:scheme}. 

A general quantity to measure the entanglement between arbitrary quantum bipartite 
systems for any mixed states is the logarithmic 
negativity~\cite{Vidal:2002,Plenio:2005,kitagawa:2006}, without the influence of 
the vacuum bath~\cite{Shapourian:2021}. In contrast, the conventional pure-state 
entanglement measures, such as the von Neumann and R\'{e}nyi entropy, capture 
both quantum and classical correlations. Since the goal of our study is harnessing 
the entanglement between the cavity and oscillator modes, we focus on the 
logarithmic negativity: $E_{N}(\rho) \equiv \log_{2}||\rho^{T_{i}}||_{1}$,
where $||X||_1=\textnormal{Tr}\sqrt{X^{\dagger}X}$ is the trace norm of the partial 
transpose $\rho^{T_i}$ with respect to the two subsystems $i=0$ (quantum-optical 
cavity mode) and $1$ (mechanical oscillator mode). The logarithmic negativity 
measures the degree to which $\rho^{T_{i}}$ fails to be positive, i.e., the extent 
of inseparability or entanglement, and it is the upper bound of the distillable 
entanglement~\cite{Vidal:2002,Plenio:2005}. The logarithmic negativity is the full 
entanglement monotone~\cite{Plenio:2005}, which satisfies the following 
criteria~\cite{vidal:2000,kitagawa:2006}: (1) $E_N$ is a non-negative functional, 
(2) $E_N$ vanishes if the state $\rho$ is separable, and (3) $E_N$ does not increase 
on average under Gaussian local operations and classical 
communication~\cite{eisert:2002,fiuravsek:2002} or positive partial transpose 
preserving operations~\cite{rains:2001}. Since $E_{N}$ quantifies the quantum 
correlation between the bipartite systems in spite of the coupling to the vacuum 
bath, the value of $E_{N}$ calculated from the cavity mode is equal to that of the 
oscillator mode: $E^{0}_{N}=E^{1}_{N}=E_{N}$, which can be verified numerically.

To characterize the quantum-entanglement control performance, we use the following 
three quantities: $\langle E_N\rangle$, $\widetilde{E}_N$, and $\widetilde{R}$ in 
open quantum optomechanical systems with either linear or nonlinear interaction 
between the quantum cavity and oscillator modes. In particular, 
$\langle E_{N}\rangle$ is the logarithmic negativity averaged over ten successive 
episodes with a single environment, $\widetilde{E}_N$ is the corresponding average 
over one episode with $T$ time steps in a single quantum environment, and 
$\widetilde{R}$ denotes the ensemble-averaged value of the reward $R$ over a small 
number of multiple parallel quantum environments for each episode. In our 
computations, all the control actions $G$, $\alpha_L$, or detuning $\Delta$, the 
nonlinear coupling $g_0$, and the dissipation coefficients $(\kappa,\gamma)$ are 
in units of $\omega_m$. The time unit is $\omega_m^{-1}$.

\subsection{RL in linear quantum optomechanics} 

A quantum optomechanical system with linear photon-phonon interactions is governed 
by the beam-splitter Hamiltonian. In an optical experimental platform, a 50:50 beam 
splitter with the transformation angle $\pi/4$ can create an entangled Bell state 
between the two input optical modes~\cite{pakniat;2017,bouchard;2020,kim;2003}. 
Similarly, in a quantum optomechanical system, Bell states between photon and phonon 
modes can be realized by controlling the beam-splitter Hamiltonian. As a result, the 
maximally attainable value of the logarithmic negativity is 
$E_{N}\sim \log 2\sim 0.7$ (in the natural logarithmic base), corresponding to the 
maximally entangled Bell state, as shown in Fig.~\ref{fig:scheme}. This ``best'' 
entangled state can be realized by the model-free PPO agent, regardless of whether 
the observation is the expectation or WCM photocurrent. To see this, we note that, 
in the beam-splitter model, the initial quantum state is set as a pure 
state~\cite{behunin:2022,wang:2015}: $|\psi\rangle=|10\rangle$, where the photon is 
in the first excited mode and the phonon is in the vacuum mode. The partial 
observable of the quantum state for the PPO agent is set as the expectation of the 
photon number $\langle\hat{n}_p\rangle(t)=\langle\hat{P}_1\rangle(t)$ or the WCM 
photocurrent 
$\sqrt{\eta}\,\mathcal{I}(t) = \langle \sqrt{\eta}\hat{P}_1\rangle(t) + \frac{dW(t)}{\sqrt{4\eta}\;dt}$.

\begin{table} [ht!]
\caption{Results of entanglement engineering from deep RL-based, Bayesian, and 
random control. The observations are the expectation of the photon number 
$\langle \hat{n}_p\rangle$ and the WCM photocurent $\mathcal{I}(t)$ at the 
measurement rate $\eta=1$. The Bayesian hyperparameter is $\lambda_{opt}=10$ for 
the $\langle \hat{n}_p\rangle$ task and $\lambda_{opt}=2$ for the $\mathcal{I}(t)$ 
task. Displayed are the results of the average logarithmic negativity 
$\langle E_{N}\rangle/\log{2}$ with the standard deviation. For training and testing 
phases, $\langle E_{N}\rangle/\log{2}$ is averaged over ten end-training or testing 
episodes, each having $T=500$ time steps. Each observation is obtained by averaging 
over $n_{\textnormal{traj}}=1$ for $\langle \hat{n}_{p}\rangle$ and 
$n_{\textnormal{traj}}=5$ for $\mathcal{I}(t)$ through simulating the SME, and 
$n_{\textnormal{traj}}$ denotes the number of independent trajectories from SME 
simulations.}
\label{ta: DRL_bayesican_random}
\begin{tabularx}{\linewidth}{YYYY}
\hline\hline
\specialrule{0em}{1pt}{1pt}
Controller & Condition &$\langle \hat{n}_p \rangle$ $n_{\textnormal{traj}}=1$ & $\mathcal{I}(t)$ $n_{\textnormal{traj}}=5$ \\
\specialrule{0em}{1pt}{1pt}
\hline
\specialrule{0em}{1pt}{1pt}
Deep RL ($\%$) & Training & $83.81\pm 1.85$ & $\mathbf{64.81\pm 1.47}$ \\
\specialrule{0em}{1pt}{1pt}
&Testing & $84.95\pm 1.99$& $\mathbf{65.01\pm 1.76}$\\
\specialrule{0em}{1pt}{1pt}
Bayesian ($\%$) & $\lambda=1$ & $56.89\pm 6.40$& $35.48\pm 5.34$\\
\specialrule{0em}{1pt}{1pt}
 & $\lambda_{opt}$ & $\mathbf{93.21\pm0.89}$& $49.24\pm0.44$\\
\specialrule{0em}{1pt}{1pt}
Random ($\%$) & & $38.15\pm9.46$ & $33.46\pm4.27$\\
\hline\hline
\end{tabularx}
\end{table}

Experimentally, directly measuring the entanglement, e.g., in terms of logarithmic 
negativity, for arbitrary entangled states is generally not viable. Identifying 
an experimentally feasible quantity to characterize the entanglement in arbitrary 
quantum systems remains challenging. We focus on the relationship between 
logarithmic negativity and the expected photon number, based on recent experiments 
on multiplexed photon number measurement~\cite{essig:2021,Thekkadath:2020,porotti:2022,johnson:2010,gleyzes:2007,puentes:2009,lvovsky:2009}. 
To proceed, we note that the beam-splitter Hamiltonian is limited to a four-level 
basis, due to the following reasons: (1) only one energy level in the cavity mode 
of the initial state has been excited from the vacuum state, i.e., 
$|\psi\rangle = |10\rangle$, (2) the linear interaction serves only to transfer 
the quantum states between the cavity and mechanical mode (i.e., no quantum 
excitation), and (3) the system couples to the vacuum bath only at absolute zero 
temperature (i.e., without any thermal excitation), thereby blocking any 
interactions between higher-level quantum states. In this case, the maximum 
logarithmic negativity $E_{N}\sim\log 2\sim 0.7$ implies that the attained quantum 
state is the following Bell state 
\begin{align} \nonumber
|\Phi^{\varphi}\rangle = \frac{1}{\sqrt{2}}\left[|10\rangle + e^{i\varphi}|01\rangle\right],
\end{align}
with the associated expected photon number 
$\langle \hat{n}^{\textnormal{target}}_p\rangle=\langle\hat{P}_1\rangle=0.5$.  
Consequently, the reward function can be set as $R_t \equiv -|O_t - 0.5|$, 
regardless of whether the observation $O_t$ is $\langle\hat{n}_p\rangle(t)$ or 
$\mathcal{I}(t)$. Because of the relatively small target value of the expected 
photon number: $\langle\hat{P}_1\rangle=0.5$, the variance $\mathcal{I}(t)$ in 
the WCM photocurrent can be reduced by a Gaussian filter~\cite{nixon:2019} with 
the weak measurement rate $\eta \leq 1$. The Gaussian kernel parameters of the 
filter such as the filter interval and the variance can be numerically chosen 
to reduce the standard deviation of the measurement photocurrent into a certain 
range, e.g., about ten times larger than the mean value (details in Subsec.~\ref{subsec:details_RL}). The PPO agent applies an updated stochastic policy to the quantum 
optomechanical environment to maximize the accumulated reward, where the action 
$G(t)$ is proportional to the amplitude of the cavity mode: 
$G(t)=g_0\bar{\alpha}_c(t)$. The action can be controlled by an incident 
laser~\cite{florian:2014} and is continuous in a certain range, e.g., 
$G \in [-5,5]\omega_m$. The decay rate of cavity and mechanical modes 
$\kappa=0.01\,\omega_m$, $\gamma=0.01\,\kappa$, respectively, because the quality 
of the mechanical oscillator mode is generally better than that of the optical 
cavity or microwave resonator mode~\cite{Aspelmeyer:2014,palomaki:2013,dobrindt:2008,shin:2022,seis:2022,liu:2024}.

\begin{figure} [ht!]
\centering
\includegraphics[width=0.9\linewidth]{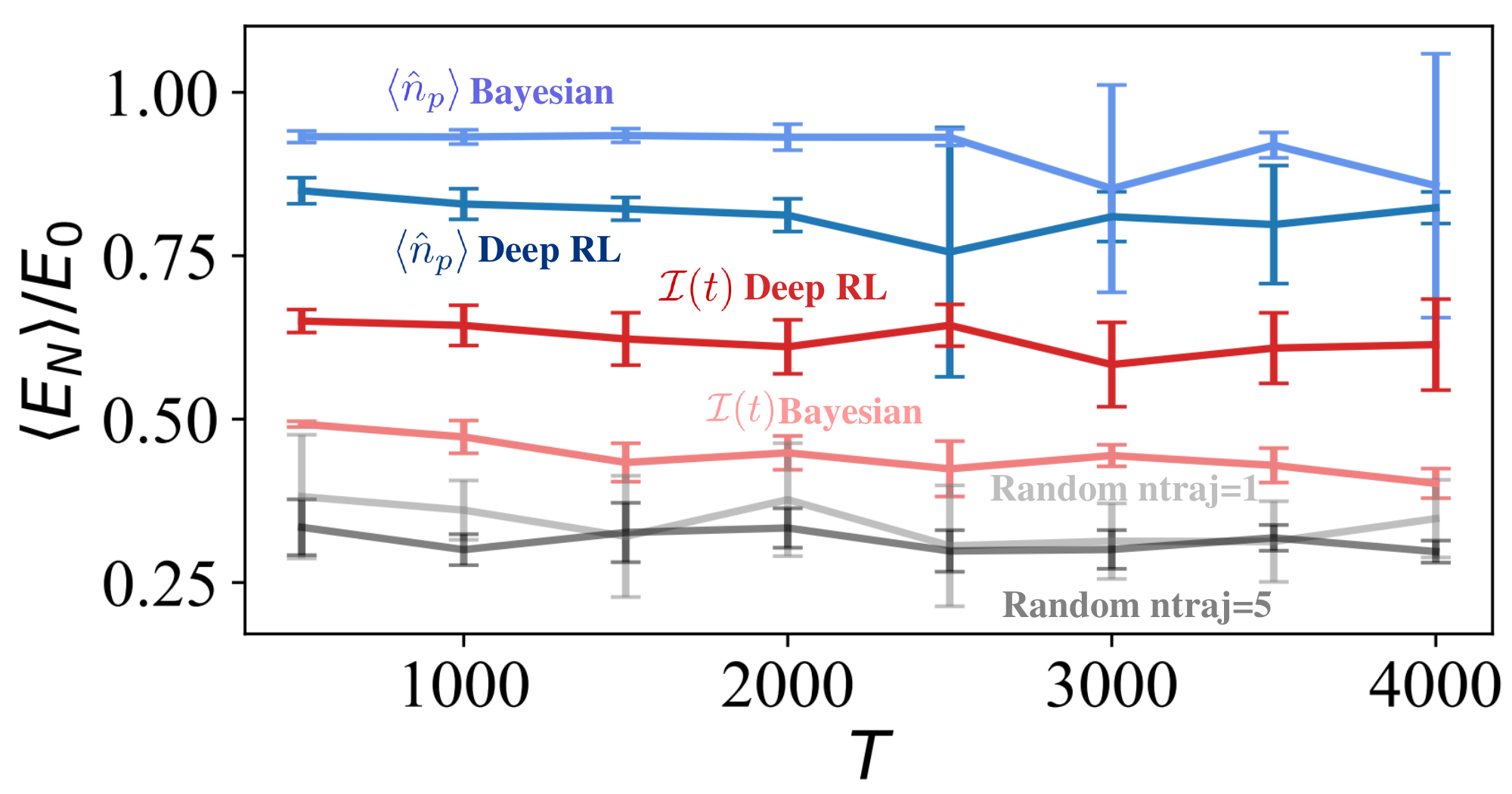}
\caption{Performance in terms of $\langle E_{N}\rangle/\log 2$ over a long time 
interval, compared for deep RL-based, Bayesian, and random control methods with 
respect to two observable options: the expected value $\langle\hat{n}_{p}\rangle$ 
and the WCM photocurrent $\mathcal{I}(t)$. The deep RL controller is trained with 
$T=500$ time steps. For all three control methods, displayed are results from the 
testing phase for the following set of time steps: 
$T=[500,1000,1500,2000,2500,3000,3500,4000]$ at the measurement rate $\eta=1$. 
The conventions, which apply to this and all subsequent figures, are as follows. 
If the vertical axis is labeled as $\langle E_N\rangle/E_0$, it represents the 
normalized logarithmic negativity, with $E_0 = \log 2\sim 0.7$ (in the natural 
logarithmic base) as the target entanglement value. Otherwise, when the vertical 
axis is labeled as $\langle E_N \rangle$, $\widetilde{E}_N$, or $E_N(t)$, it 
represents the original value of the logarithmic negativity.}
\label{fig: T_test}
\end{figure}

\begin{figure*} [ht!]
\centering
\includegraphics[width=0.9\linewidth]{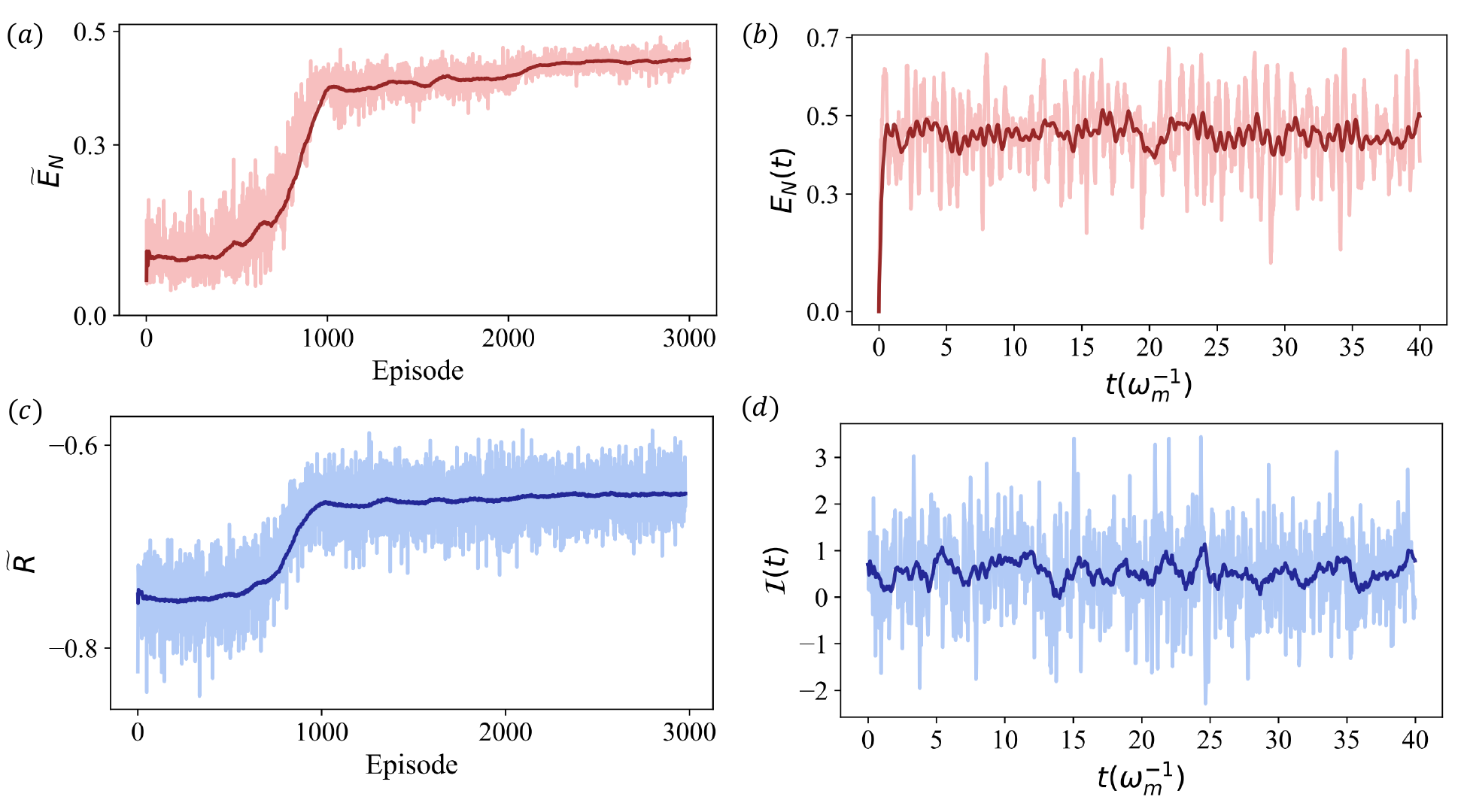}
\caption{Performance of deep-RL agent in the online training and testing phase.
The characterizing quantities are the logarithmic negativity $E_{N}$ and the 
reward function $R$ with the measurement rate $\eta=1$. (a,c) Performance 
measures in the online training phase, where the mean $\widetilde{E}_{N}$ is over 
one episode with $T=500$ time steps on the fifth quantum environment (only one 
environment) and the mean reward $ \widetilde{R}$ is obtained from $\mathbb{N}=5$ 
parallel quantum environments.  (b,d) Performance measures during the testing phase, 
where the logarithmic negativity $E_{N}(t)$ and WCM photocurrent $\mathcal{I}(t)$ 
are obtained with $T=4000$ time steps. The solid traces represent the moving-window 
average over $100$ episodes for (a,c) and $100$ time steps for (b,d).}
\label{fig: measureRate_training_testing}
\end{figure*}

Our deep RL, a model-free learning method, is implemented in the measurement-based 
feedback control framework for entanglement engineering in open quantum 
optomechanics. Details about the PPO algorithm applied in the linear quantum 
optomechanics are presented in Appendix~\ref{sec:appendix_RL_linear}. To appreciate 
its performance, we employ two benchmark methods for comparison: 
Bayesian~\cite{Stockton:2004,Borah:2021} and random control. Bayesian 
control~\cite{Stockton:2004, Borah:2021} is a state-based feedback control of the 
stochastic process as governed by the SME. In our case, the control law is given
by $G(t) = -\lambda|\langle \hat{n}_p\rangle(t) - 0.5|\omega_{m}$ 
with $\langle \hat{n}_p\rangle(t)$ being the observation, where the hyperparameter 
$\lambda$ can be numerically optimized based on the performance. If the observation 
is $\mathcal{I}(t)$, the control flow will be in the form 
$G(t) = -\lambda|\mathcal{I}(t) - 0.5|\omega_m$, in which the Wiener process blocks 
the performance to some degree. In Bayesian control, the smaller the variance in 
the measured photocurrent, the better the performance. For the random control 
method, the flow is generated by a uniform distribution in the action range 
$G\in[-5,5]\,\omega_m$. Note that the actions $G$ of random control and deep RL 
are in the same range while the one of Bayesian control is determined by the 
hyperparameter $\lambda$ and the state-based observation value or the WCM 
photocurrent. To make a fair comparison, $\lambda_{opt}$ is optimized within the 
action range $G\in[-5,5]\omega_{m}$. Specifically, the optimized hyperparameter 
$\lambda_{opt}$ corresponds to the best performance of Bayesian control in the set 
$\lambda\in\{1,2,\ldots,\lambda_{max}\}$, where $\lambda_{max}$ is the maximum 
integer of $\lambda$ to guarantee the action range $G\in[-5,5]\,\omega_m$.

Table~\ref{ta: DRL_bayesican_random} displays the values of the averaged logarithmic 
negativity $\langle E_{N}\rangle/\log 2$ from the deep RL, Bayesian, and random 
control methods. From the SME simulations, when the observation is the expectation 
of the photon number, the Bayesian control with the optimized hyperparameter 
outperforms the deep RL method. However, when the observation is the WCM 
photocurrent, the deep RL control outperforms the Bayesian method. This is promising 
as the WCM photocurrent is directly experimentally accessible while the expected 
photon number is not. Regardless of the observation, random control is generally 
ineffective. The results by deep RL control from the observation of WCM photocurrent 
tend to reduce the performance by about $20\%$ compared to that based on the 
expected photon number. For Bayesian control, the reduction is about $40\%$. 
Moreover, Fig.~\ref{fig: T_test} compares the long-time entanglement engineering 
for three control methods. Especially, for deep RL control, the PPO agent is 
trained with $T=500$ time steps but tested with a longer time horizon, e.g., 
$T=4000$ steps, including the unexplored regime by the PPO agent. It is worth 
noting from Fig.~\ref{fig: T_test} that the performance of deep RL with the 
observation of WCM photocurrent exhibits a more stable and smaller variance 
compared to the case where the observation is the expected photon number, 
especially after $T=2000$. Overall, with the experimentally feasible observation 
of WCM, the deep RL controller stands out as the choice of entanglement control 
for quantum optomechanical systems. 

\begingroup
\makeatletter
\renewcommand{\fnum@figure}{\textcolor{blue}{\figurename~\thefigure}} 
\begin{figure}[ht!]
\centering
\includegraphics[width=0.9\linewidth]{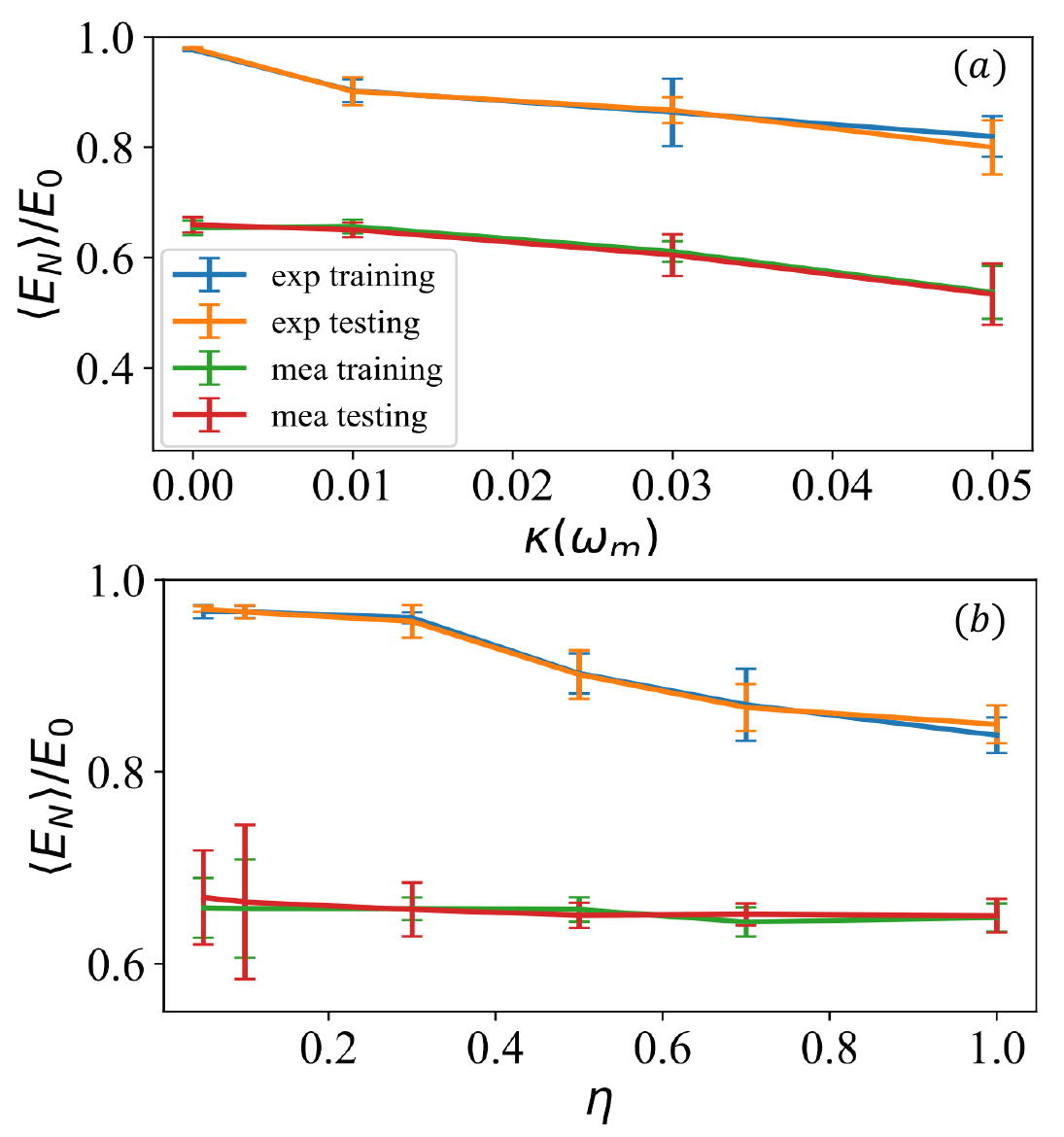}
\caption{Effects of decay and measurement rates on the control performance. Shown are the values of the average logarithmic negativity for (a) decay rates 
$\kappa=[0,0.01,0.03,0.05]\,\omega_m$ with $\eta=0.5$ and $\gamma=0.01\,\kappa$, 
and (b) measurement rates $\eta=[0.05,0.1,0.3,0.5,0.7,1]$ with 
$\kappa=0.01\,\omega_m$ and $\gamma=0.01\,\kappa$. The error bars represent the 
standard deviation of the data points. The average operation is over ten 
end-training or testing episodes. The training and testing time steps are the 
same: $T=500$.}
\label{fig: mea_exp_kappa_eta_plot}
\end{figure}
\makeatother
\endgroup

We characterize the performance of our deep-RL-based control method in terms of 
the dissipation rate, measurement rate, and the randomness effect for the initial 
state. For the measurement rate $\eta=1$, the PPO agent is sequence-wise trained 
with the WCM photocurrent. Figures~\ref{fig: measureRate_training_testing}(a) and 
\ref{fig: measureRate_training_testing}(c) show the average logarithmic negativity 
$\widetilde{E}_N$ and the mean reward $\widetilde{R}$, respectively, versus the 
episode during the training phase, in which $\widetilde{E}_N$ and $\widetilde{R}$ 
are averaged over one and five parallel quantum environments, respectively. Both 
quantities ultimately converge due to the properly designed reward function 
$R(t)=-|\mathcal{I}(t)-0.5|$. Note that the variance of $\widetilde{E}_N$ is 
suppressed with the episodes, implying the mixture-robust nature of entanglement 
in the quantum optomechanical system. The testing phase is longer ($T=4000$ time 
steps) than the training phase ($T=500$ time steps) and the corresponding 
performance measures are shown in Figs.~\ref{fig: measureRate_training_testing}(b) 
and \ref{fig: measureRate_training_testing}(d). In addition to the variance in the 
learning of the deep RL agent with the stochastic policy, the Gaussian Wiener 
process in the WCM photocurrent and the stochastic collapse process stipulated by 
the SME also contribute to the variances of the performance measures. However, the 
deep RL control still manages to maintain the solid traces of the testing 
$\mathcal{I}(t)$ around the target value 
$\langle \hat{n}^{\textnormal{target}}_p\rangle=0.5$ in 
Fig.~\ref{fig: measureRate_training_testing}(d) and the resulting entanglement 
quantity $E_N(t)$ is displayed in Fig.~\ref{fig: measureRate_training_testing}(b).

Since the quantum optomechanical system is coupled to the vacuum bath, the coupling 
strength or disturbance between the classical and quantum environments will affect 
the control performance, as exemplified in Fig.~\ref{fig: mea_exp_kappa_eta_plot}(a).
Previous experiments~\cite{Aspelmeyer:2014,song:2014,dobrindt:2008,palomaki:2013} 
demonstrated that the quality of the mechanical oscillator is generally better than 
that of the optical cavity or microwave resonator, i.e., $\gamma<\kappa$, so we set 
the decay rate of the oscillator at two orders of magnitude smaller than that of 
the cavity~\cite{qian:2012}: $\gamma=0.01\,\kappa$. 
Figure~\ref{fig: mea_exp_kappa_eta_plot}(a) shows, for both the expectation and 
the measurement flow observations, the performances of the training and testing 
processes, which are consistent with each other in the sense that their mean values 
decrease and the variances increase with the decay rate. The origin of the 
performance fluctuations is the classical dissipation to the vacuum bath, rendering 
the system less controllable by laser.

The uncertainty in the classical information extracted from the quantum system 
depends on the discrete-time step size $dt$ and the measurement rate $\eta$, which 
directly determines the degree of the quantum-state stochastic collapse and quantum 
decoherence from the WCM term in the SME. If the expectation of the photon number 
is the observation, the stronger the measurement rate (proportional to the 
measurement strength), the poorer the performance of deep-RL control as characterized
by a decrease in the mean values and an increase in the uncertainties of $E_{N}$, 
as shown in Fig.~\ref{fig: mea_exp_kappa_eta_plot}(b), which originate from the 
intrinsic random process in the SME induced by the measurement process. However, 
if the observation is the WCM photocurrent, the weaker measurement rate will 
introduce larger variances in the observation signal and reduce the stochasticity
of the process due to the incomplete/partial extracted information as described by 
the SME. In our case, the target mean value, 
$\langle \hat{n}^{\textnormal{target}}_p\rangle=0.5$, is on the order of $10^{-1}$, 
rendering necessary introducing a Gaussian filter to reduce the uncertainty. The 
resulting performance of deep-RL control is approximately the same for 
$\eta\in[0.05,1]$, as shown in Fig.~\ref{fig: mea_exp_kappa_eta_plot}(b).

\begin{figure} [ht!]
\centering
\includegraphics[width=0.9\linewidth]{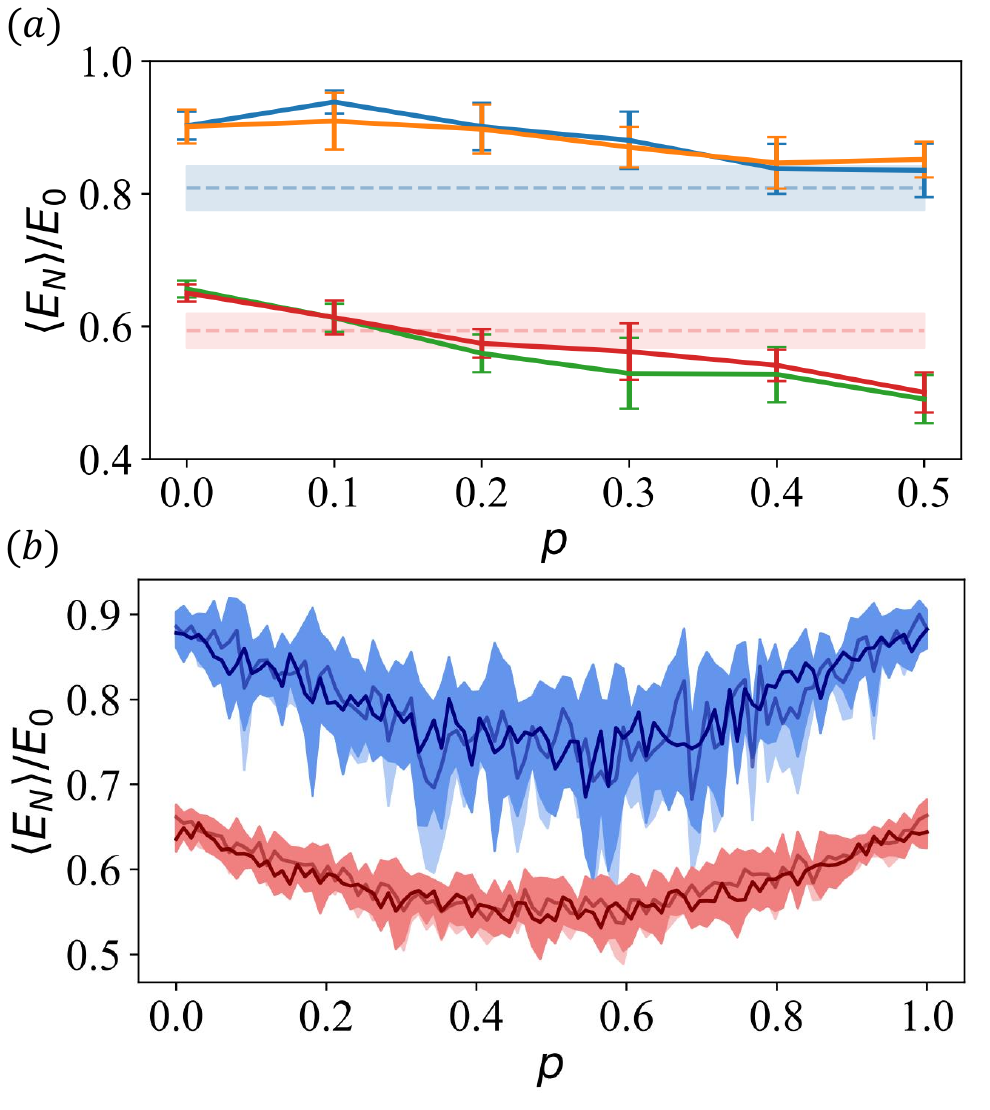}
\caption{Robustness of deep-RL method trained with pure or mixed states. (a) In the 
training and testing phase, performance of $\langle E_{N}\rangle/E_0$ for different 
initial mixed states (solid traces): 
$\rho=(1-p)|10\rangle\langle 10|+ p|01\rangle\langle 01|$ with 
$p=[0,0.1,0.2,0.3,0.4,0.5]$. The dashed traces indicate the performances trained 
with random initial mixed states with the random variable $p\in[0,0.5]$. (b) Testing
performance of two kinds of trained agents with $p\in[0,1]$: one trained with the 
pure initial state $|\psi\rangle=|10\rangle$ and another with random initial mixed 
states, which are distinguished by the color depth of the curve and the error bars. 
The blue and red curves denote the performances with the observation 
$\langle \hat{n}_p\rangle$ and $\mathcal{I}(t)$, respectively, with error bars. 
The measurement rate is $\eta=0.5$, and the training and testing time steps are 
$T=500$.}
\label{fig: mea_exp_random_initial}
\end{figure}

\begin{figure*} [ht!]
\centering
\includegraphics[width=0.9\linewidth]{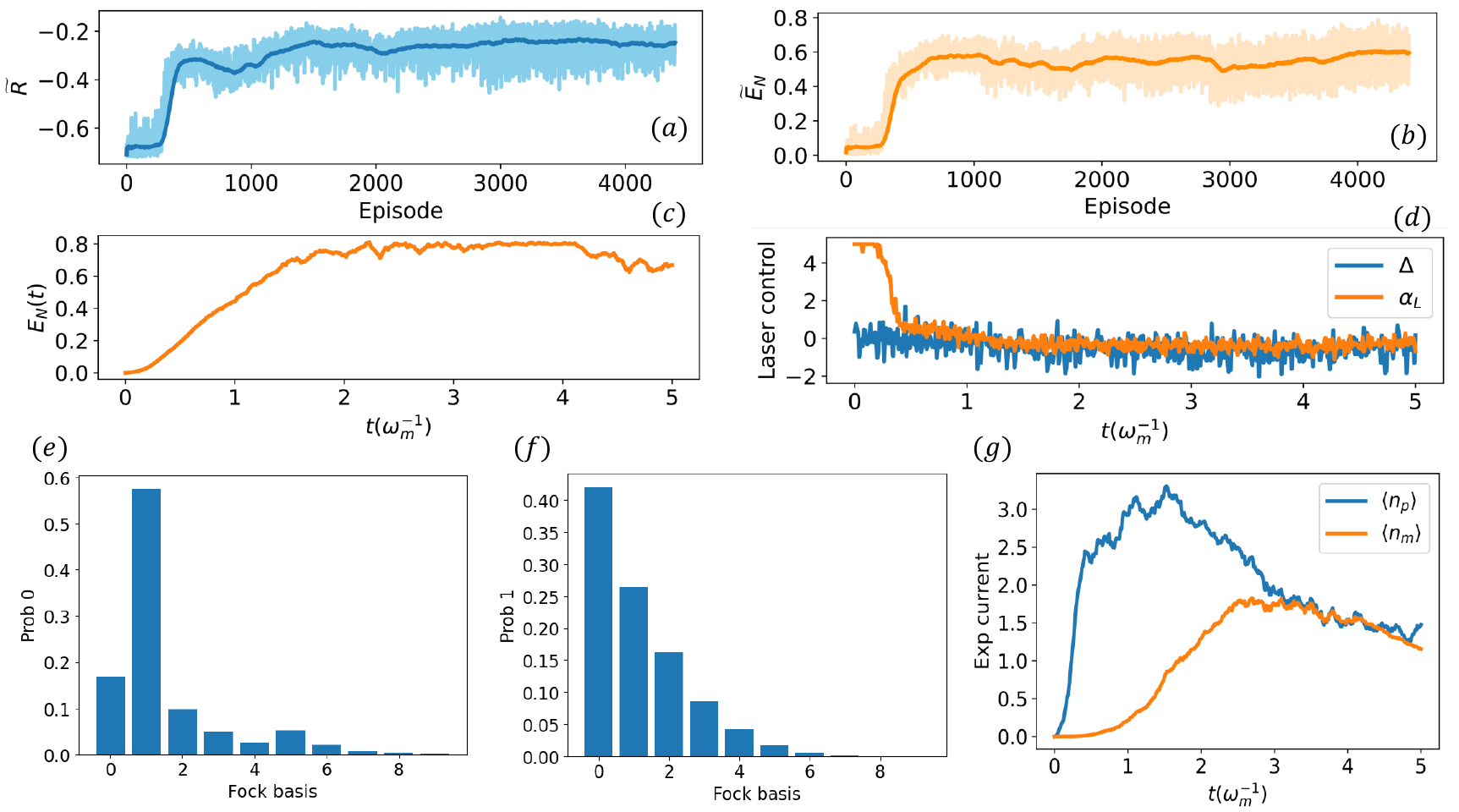}
\caption{Generating target for deep-RL based creation and stabilization of 
entanglement in a nonlinear open quantum optomechanical system. (a,b) Trained 
quantities $\widetilde{R}$ and $\widetilde{E}_N$ converge to a certain value as 
the episode number increases, as illustrated by the light-color curves, where the 
dark blue and orange traces represent the data averaged over 100 previously 
consecutive episodes. (c,d) Time-dependent series of $E_{N}(t)$ and the driven 
laser signals $\Delta,\alpha_L$ at a certain episode selected from the training 
converged regime in (a,b). (e,f) The corresponding photon and phonon statistics 
on the Fock basis at the end of the time point of the selected training episode 
in (c,d). (g) The time evolution of the corresponding expected quantities, 
including the expected numbers $\langle \hat{n}_p\rangle$ and 
$\langle \hat{n}_m\rangle$ in the Fock basis, where the time series of 
$\langle \hat{n}_p\rangle$(t) serves as the target to construct the reward function 
in the next phase, i.e., the experimental version shown in 
Fig.~\ref{fig: WCM_nonlinear_obser_Pn}.}
\label{fig: WCM_nonlinear_obser_EN}
\end{figure*}

Experimentally, mixed quantum states are more realizable than pure states due to 
the quantum decoherence with the classical environment, e.g., the vacuum bath. 
To address this issue, and referring to the previous work~\cite{Yuan:2020}, we 
assume that the initial state is a mixed state in the form of 
$\rho=(1-p)|10\rangle\langle 10|+ p|01\rangle\langle 01|$, where the parameter $p$ 
is fixed or a random variable $p\in[0,1]$ because of the coupling to the classical 
environment. The beam-splitter Hamiltonian stipulates that the photon and phonon 
modes are symmetric to each other, allowing $p$ to be rescaled to the interval
$p\in[0,0.5]$. Figure~\ref{fig: mea_exp_random_initial}(a) shows the performance 
with respect to the initial mixed quantum state with the same parameter $p$ for 
each training and testing episode (solid traces), where the complete mixed case 
with $p=0.5$ leads to the worst performance but still possesses entanglement to
a significant extent. The reason lies in the inherent property of the beam-splitter 
Hamiltonian, which can create the maximum entangled states: 
$[|10\rangle+e^{i\varphi}|01\rangle]/\sqrt{2}$, with respect to the part of the 
initial quantum state, such as $|10\rangle$ or $|01\rangle$ through the linear 
interactions, regardless of whether it acts on a pure or a mixed state. In 
Fig.~\ref{fig: mea_exp_random_initial} (a), the dashed traces display the 
performance during the training phase with a random initial mixed quantum state, 
which is generated by the random variable $p$ with the uniform distribution in the 
range of $p\in[0,0.5]$. The error bar characterizes the uncertainty over ten 
end-training episodes. 

Figure~\ref{fig: mea_exp_random_initial}(b) shows the testing performance of two 
kinds of trained models, one trained by the initial state $|\psi\rangle=|10\rangle$ 
and another by the random initial mixed-state 
$\rho=(1-p)|10\rangle\langle 10|+ p|01\rangle\langle 01|$ (distinguished by dark 
and light colors, respectively). Note that the beam-splitter Hamiltonian transforms 
the initial state $|10\rangle$ or $|01\rangle$ to a Bell state with the corresponding
expected photon number: $\langle\hat{n}^{\textnormal{target}}_p\rangle=0.5$, where 
the dissipative degree to the vacuum bath is much weaker than the beam-splitter 
interaction. However, if the initial state is the mixed state, the 
$|10\rangle\langle 10|$ and $|01\rangle\langle 01|$ components will become
independently entangled, resulting in the total quantum state being a mixture of 
two entangled Bell states. As a result, a nontrivial entanglement value is expected 
for the initial mixed state governed by the beam-splitter Hamiltonian. With the 
mixed probability $p=0.5$, it results in an equal mixture of the Bell states, as 
shown in Fig.~\ref{fig: mea_exp_random_initial}. In the testing phase, the two 
trained models use the same initial state for a fixed value of $p$. The two models 
have a comparable performance, suggesting that the deep RL method is robust to the 
initial randomness in a mixed state. More specifically, during the testing phase, 
the observation is the expected photon number or WCM photocurrent. The worse 
performance occurs for $p=0.5$ and for other values of $p$, the performance is 
symmetric about $p=0.5$ due to symmetric role of the photon and photon modes in 
the beam splitter Hamiltonian. Note that the model trained with the observation 
being the measurement photocurrent displays a small difference in the performance 
measure [$\langle E_{N}\rangle/E_0$ over the whole probability interval $p\in[0,1]$] 
between the best and worst cases, with less uncertainties than the case where the 
observation is the expected photon number. Taken together, our deep-RL model 
trained by the weak measurement photocurrent holds a lower mean performance but 
possesses robustness against mixed quantum states compared with the scenario based 
on observing the expected value of the photon number, due to the strong capability 
of RL in learning randomness and executing accurate high-dimensional data-fitting.

\subsection{RL in nonlinear quantum optomechanics}

\begin{figure*} [ht!]
\centering
\includegraphics[width=0.9\linewidth]{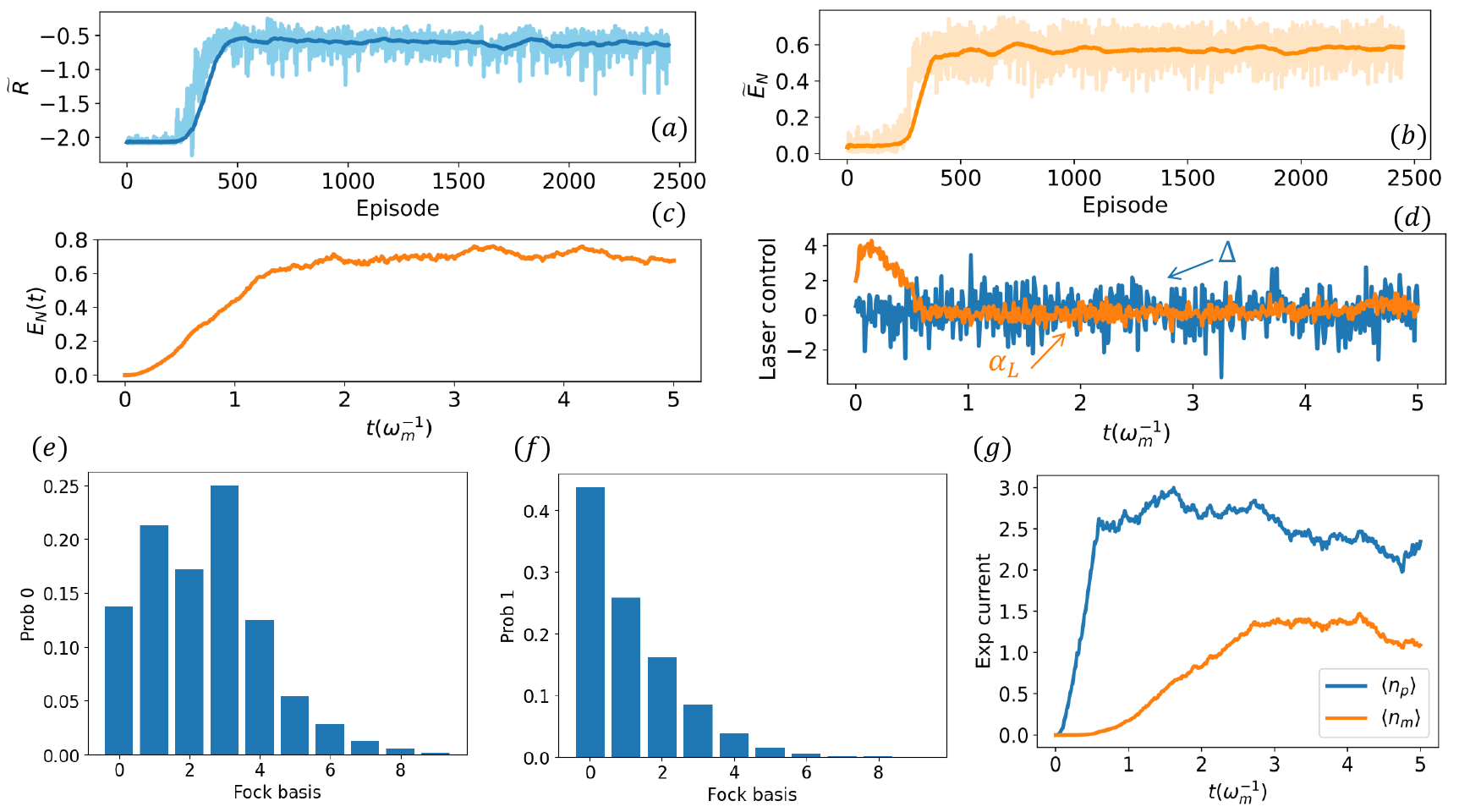}
\caption{Entanglement engineering by the recurrent PPO agent. The target 
generated as described in Fig.~\ref{fig: WCM_nonlinear_obser_EN} is exploited to 
create entanglement by $E_{N}\sim\log2$ from the only partial observation of the 
expected photon number $\langle \hat{n}_p\rangle(t)$. The reward function is 
$R(t)=-|\langle \hat{n}_p\rangle(t)-\langle \hat{n}_p^{\textnormal{target}}\rangle(t)|$, where the target time series $\langle \hat{n}_p^{\textnormal{target}}\rangle(t)$ 
is from the target-generating process in Fig.~\ref{fig: WCM_nonlinear_obser_EN}(g). 
In this training configuration, while only partial information is extracted from the 
system, the performance measures in (a-g) display a similar behavior compared with 
those in Fig.~\ref{fig: WCM_nonlinear_obser_EN}. Other aspects of the setting and 
parameters are the same as in Fig.~\ref{fig: WCM_nonlinear_obser_EN}.}
\label{fig: WCM_nonlinear_obser_Pn}
\end{figure*}

In an open quantum optomechanical system under the strong laser-driven approximation, 
the radiation pressure on the movable mechanical mirror generates a linear 
interaction between the optical and mechanical modes. When this approximation does 
not hold, the interaction between the two modes becomes nonlinear. Entanglement can 
still be created despite the nonlinear interaction, but control becomes more 
challenging. In particular, in the standard quantum optomechanical system, the 
nonlinear coupling term $\hbar g_0\hat{a}^{\dagger}\hat{a}(\hat{b}^{\dagger}+\hat{b})$ 
can be used to create entanglement, but high-level quantum states can also be excited 
during the process, making it difficult to stabilize the entanglement within a finite 
Fock basis. Realistically, the quantum dynamics are governed by the SME due to the 
WCM, which induces the nonlinear stochastic evolution. The problem then becomes that 
of creating and stabilizing the entanglement of non-Gaussian states decaying to the 
vacuum bath. Despite the difficulties, model-free deep-RL can still provide a general approach through some optimal combination of the neural network structure, observable, 
reward function, and action.

We consider the nonlinear optomechanical system and exploit deep RL to set the 
control goal of achieving the entanglement near $E_{N}\sim\log2$. This nonlinear 
entangled state shares a similar entanglement value with the maximum entangled Bell 
state in the corresponding linear system. For entanglement engineering of a nonlinear 
optomechanical system, a key issue is selecting an effective and experimentally 
feasible observation quantity. Utilizing a general actor and critic neural network, 
the deep RL agent can learn the relationship between entanglement and the experimental
observables of the optomechanical system in a model-free manner. To achieve control, 
we articulate a training process consisting of two phases: the target-generating 
phase and the target-utilization phase, facilitated by deep RL. 

The first training step is the target-generating phase, in which numerical SME 
simulations are used to generate the observation and reward data and the PPO 
agent interacts with the quantum environment, observes the logarithmic negativity 
$E_{N}(t)$ and constructs the reward function combining the expectation number 
of the photons and phonons: 
$R(t)=-|E_{N}(t)-\log2|-|\langle \hat{n}_p\rangle(t)+\langle \hat{n}_m\rangle(t)-a|/b$ 
with numerically optimized hyperparameters $a=1$ and $b=30$. (Note that direct 
experimental measurement of the logarithmic negativity is currently not available.) 
Figure~\ref{fig: WCM_nonlinear_obser_EN} shows the control results, where the 
excitation of quantum states is limited by the total number 
$\langle \hat{n}_p\rangle+\langle \hat{n}_m\rangle$. The target time series of the 
expected photon number is obtained as 
$\langle \hat{n}_p^{\textnormal{target}}\rangle(t)$. The second step is the 
target-utilization phase, during which the reward function is 
$R(t)=-|\langle\hat{n}_p\rangle(t)-\langle\hat{n}_p^{\textnormal{target}}\rangle(t)|$.

Since it is time-dependent, the recurrent neural network added after the MLPs in 
the PPO agent displays a strong and stable learning ability, which outperforms the 
case with only MLPs. The expected photon number $\langle \hat{n}_p\rangle(t)$ is 
observed by the recurrent PPO agent as 
$\langle \hat{n}_p\rangle=\sum_{n}n\langle\hat{P}_n\rangle$, which is experimentally 
more feasible than the quantity $E_{N}$. While the recurrent neural network has some 
considerable advantages, such as long-term momery~\cite{hochreiter:1997}, it still 
encounters the challenge of engineering optimization~\cite{pleines:2022} in order 
to achieve a correct and efficient implementation. In our case, the main challenge 
is the time cost to optimize the parameters to search for a global minimum or maximum 
due to the ten stochastic collapse operators, $\hat{P}_n=|n\rangle\langle n|$ with 
the respective Fock numbers $n=0,1,\ldots,9$, in the SME with the measurement rate 
$\eta=0.1$, requiring a long simulation time. Our solution is to consider only the 
$\mathbb{N}=1$ quantum optomechanical environment, in which the agent collects data 
and updates the policy every $\mathbb{Z}=15$ and $\mathbb{Z}=5$ episodes in two 
phases (target-generating and target-utilization), respectively, with the time 
horizon $T=500$. Note that, using ten stochastic projectors $\hat{P}_n$ can result 
in a large variance in the WCM photocurrent:
\begin{align} \nonumber
\sqrt{\eta}\,\mathcal{I}(t)=\sum_{n}n\left[\langle\sqrt{\eta}\hat{P}_n\rangle+\frac{dW_n(t)}{\sqrt{4\eta}dt}\right],
\end{align}
where ten independent Wiener processes $dW_n(t)$ are used. In this case, observation 
of the measured random photocurrent is infeasible. Even if the deep RL agent is 
trained in two phases with the expected photon number, it can fail during the 
training process due to the numerical cutoff in the Hilbert space dimension and 
the strong randomness introduced by the SME. In the nonlinear quantum optomechanical 
system, the interaction strength is $g_0=0.2\,\omega_m$. The PPO agent creates 
entanglement characterized by $E_{N}\sim \log2$ versus time, calculated through 
the SME with dissipation to the vacuum bath for $\kappa=0.1\,\omega_m$, and 
$\gamma = 0.01\,\kappa$. The system is initialized in the vacuum state 
$|\psi\rangle = |00\rangle$, i.e., the pure state, with $10\times 10$ Fock bases. 
The time-dependent control signal is the detuning $\Delta$ and the amplitude 
$\alpha_L$ of the driven laser within the fixed range 
$\Delta,\alpha_L\in[-5,5]\,\omega_m$.

\begin{figure} [ht!]
\centering
\includegraphics[width=0.9\linewidth]{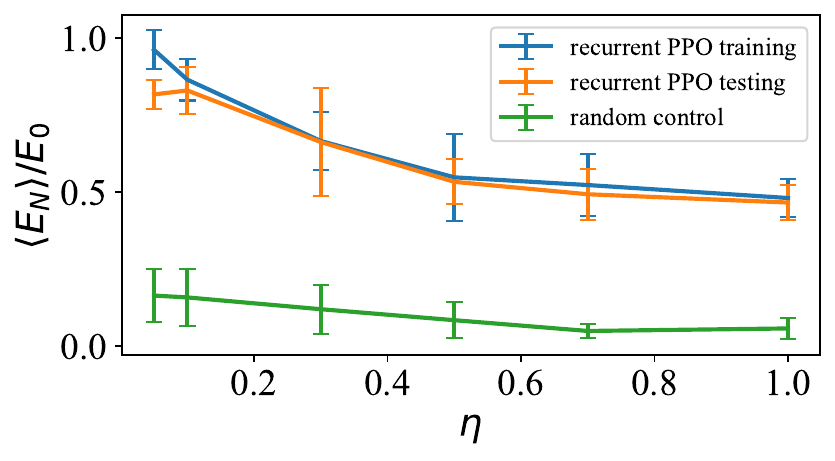}
\caption{Target-utilization phase of entanglement engineering of a nonlinear
optomechanical systems. Shown are the results of online training and testing 
of the entanglement measure $\langle E_{N}\rangle/E_0$ for measurement rates
$\eta=[0.05,0.1,0.3,0.5,0.7,1]$, in comparison with the benchmark performance 
of random control. The error bars are the corresponding standard deviation.
The results from random control flow are also included for comparison. Other 
parameters are the same as those in Fig.~\ref{fig: WCM_nonlinear_obser_EN}.}
\label{fig: performance_nonlinear_PPO_random}
\end{figure}

Representative results are as follows. In the target-generating phase, despite the 
disturbance of the stochastic process from WCM, the training curves for both the 
reward $\widetilde{R}$ and the logarithmic negativity $\widetilde{E}_N$ converge 
with the episode number, as shown in Figs.~\ref{fig: WCM_nonlinear_obser_EN}(a,b), 
indicating that entanglement has been created and stabilized by the well-trained 
PPO agent, as shown by Fig.~\ref{fig: WCM_nonlinear_obser_EN}(c) with the laser 
control signal displayed in Fig.~\ref{fig: WCM_nonlinear_obser_EN}(d). At the end 
of the time period, the photon and phonon statistics with respect to the Fock basis 
are shown in Figs.~\ref{fig: WCM_nonlinear_obser_EN}(e,f), where the reduced 
photon state exhibits an oscillating tail that resembles the displaced squeezed 
state and the reduced phonon state displays the thermal-like state. 
Figure~\ref{fig: WCM_nonlinear_obser_EN}(g) shows the corresponding target pattern 
$\langle \hat{n}^{\textnormal{target}}_p\rangle(t)$. In the target-utilization phase, 
the recurrent PPO agent is able to steadily learn to create and stabilize the 
entanglement, as shown in Fig.~\ref{fig: WCM_nonlinear_obser_Pn}, where only 
partial information is extracted from the quantum optomechanical environment. 
Especially, various entangled states have been created, such as a reduced photon 
state with the head oscillating on the Fock basis in photon statistics entangled 
with the thermal-like reduced phonon state, as exemplified in 
Figs.~\ref{fig: WCM_nonlinear_obser_Pn}(e,f). Due to the nonlinear and stochastic
process in the SME, the entangled states created and controlled are not steady
states, rendering infeasible Bayesian control. We thus employ random control as a
benchmark, where a uniformly random distribution of actions is taken in a certain 
range $\Delta,\alpha_L\in[-5,5]\omega_m$ and the tested values of the measurement 
rate are $\eta=[0.05,0.1,0.3,0.5,0.7,1]$. 
Figure~\ref{fig: performance_nonlinear_PPO_random} shows that, as the measurement 
rate increases, the random control is unable to harness the entanglement while our 
well-trained recurrent PPO agent can maintain the entanglement percentage at $50\%$ 
or higher. 

\subsection{Physical understanding of entanglement engineering through model-free deep RL}

In an experiment, it is usually difficult to directly obtain information about the 
entanglement. For entanglement engineering of a quantum optomechanical system, one 
scenario is that the RL agent observes the photon number to steer the laser to create 
and stabilize entanglement, as illustrated in Fig.~\ref{fig: sche_phy_under}. Here
we provide a physical interpretation of RL control for entanglement engineering in 
both the linear and the nonlinear interaction regimes. The key physical relationships
involved are that between the entanglement and photon number, and that between the 
photon number and laser driving. We also describe the capability of the RL agent to 
train the laser driving to modulate the two-mode interaction to reduce quantum 
decoherence resulting from WCM and the quantum dissipation to the vacuum bath.

\subsubsection{Linear interaction regime} 

\begin{figure} [ht!]
\centering
\includegraphics[width=0.9\linewidth]{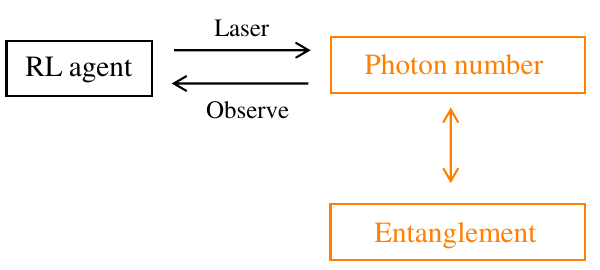}
\caption{RL-based entanglement engineering of a quantum optomechanical system.}
\label{fig: sche_phy_under}
\end{figure}

For the linear quantum optomechanical system, the maximum entanglement corresponds 
to a Bell state, of which the expected photon number is $\langle\hat{n}_p\rangle=0.5$.
Intrinsically, the beam splitter Hamiltonian is capable of generating Bell 
states~\cite{pakniat;2017,bouchard;2020,kim;2003}, a reasonable assumption is that, 
when the expected photon number reaches the value of 0.5, the maximum entanglement 
is achieved in a linearly interacting quantum optomechanical system. This assumption 
provides the base for constructing the reward function 
$R(t)=-|\langle \hat{n}_p\rangle(t)-0.5|$, where the deviation in the expected photon 
number from 0.5 results in a decreasing reward and therefore implies reduced 
entanglement. As illustrated in Fig.~\ref{fig: sche_phy_under}, the RL agent is 
designed to maximize the accumulated reward value, which is equivalent to stabilizing 
the expected photon number about the value of 0.5 for as long as possible. The 
testing results shown in Fig.~\ref{fig: measureRate_training_testing} indicate that 
the maximum entanglement can indeed be created and stabilized by the RL control. 

A central step in RL control is to modulate the laser input based on the measured 
photon number, which requires the relationship between the laser driving and the 
photon number. When the frequency of the laser is in the red-detuned regime: 
$\Delta = \omega_L - \omega_c = -\omega_m$, the quantum state switches between the 
two modes - the cavity optical and the mechanical oscillator modes, leading to 
a ``swap'' Hamiltonian. The coefficient $G$ is proportional to the amplitude of the 
cavity parameter $\bar{\alpha}_c$ that is determined by the laser. In the linear 
interaction regime, RL control is achieved via two adjustments of the laser based 
on the measured photon number: (1) the laser frequency is changed into the 
red-detuned regime and (2) the laser amplitude is perturbed to modulate the driving 
strength $G$ to control the two modes of switching, which affects the expected photon 
number. Note that, during this process, there is no energy gain: there is energy loss 
due to the dissipation of the cavity and oscillator modes into the vacuum bath with 
the dissipation rate given by $\gamma=0.01\,\kappa$. This relation means that the 
energy loss due to the oscillator mode occurs more slowly than that with the cavity 
mode. In essence, the working of the laser is to transfer the energy from the 
oscillator mode to the cavity mode to stabilize the photon number to a desired value. 
The underlying dissipation process is not beneficial to the entanglement, as it cannot
be modulated by the ``swap'' term in the Hamiltonian, eliminating any possibility of 
entanglement enhancement in an optomechanical system in the linear interaction regime.
It is worth noting that, in the nonlinear interaction regime, entanglement enhancement
and dissipation reduction are possible, as will be described below.

\subsubsection{Nonlinear interaction regime}

When the interactions between the optical and mechanical modes are nonlinear, the 
relationship between entanglement and photon number can be sophisticated and is 
currently unknown. However, model-free deep RL can be used to find the relation 
numerically. To achieve this, we first assume that there is a solution of the 
one-to-one correspondence between $E_N$ and $\langle \hat{n}_p \rangle$ in the time 
domain. The reward function is constructed according to the target entanglement 
$E_N =\log 2$ to train the RL agent to maximize the accumulated reward. In the 
testing phase, the time-dependent series of the expected photon number controlled 
by the well-trained PPO in Fig.~\ref{fig: WCM_nonlinear_obser_EN}(g) is regarded 
as the target time series of the expected photon number for the next 
target-utilization phase. Note that the ``best'' photon number is no longer simply 
0.5: it is now time-dependent. In the next training phase, the reset RL agent will 
learn to control the system with the observation $\langle\hat{n}_p\rangle(t)$ based 
on the target's expected photon number 
$\langle\hat{n}^{\textnormal{target}}_p\rangle(t)$. The performance of the new RL 
agent in the testing phase, as shown in Fig.~\ref{fig: WCM_nonlinear_obser_Pn}, 
validates our initial assumption about the existence of the one-to-one correspondence 
between $E_N$ and $\langle \hat{n}_p\rangle$, even though it is time-dependent. 

\begin{figure} [ht!]
\centering
\includegraphics[width=0.9\linewidth]{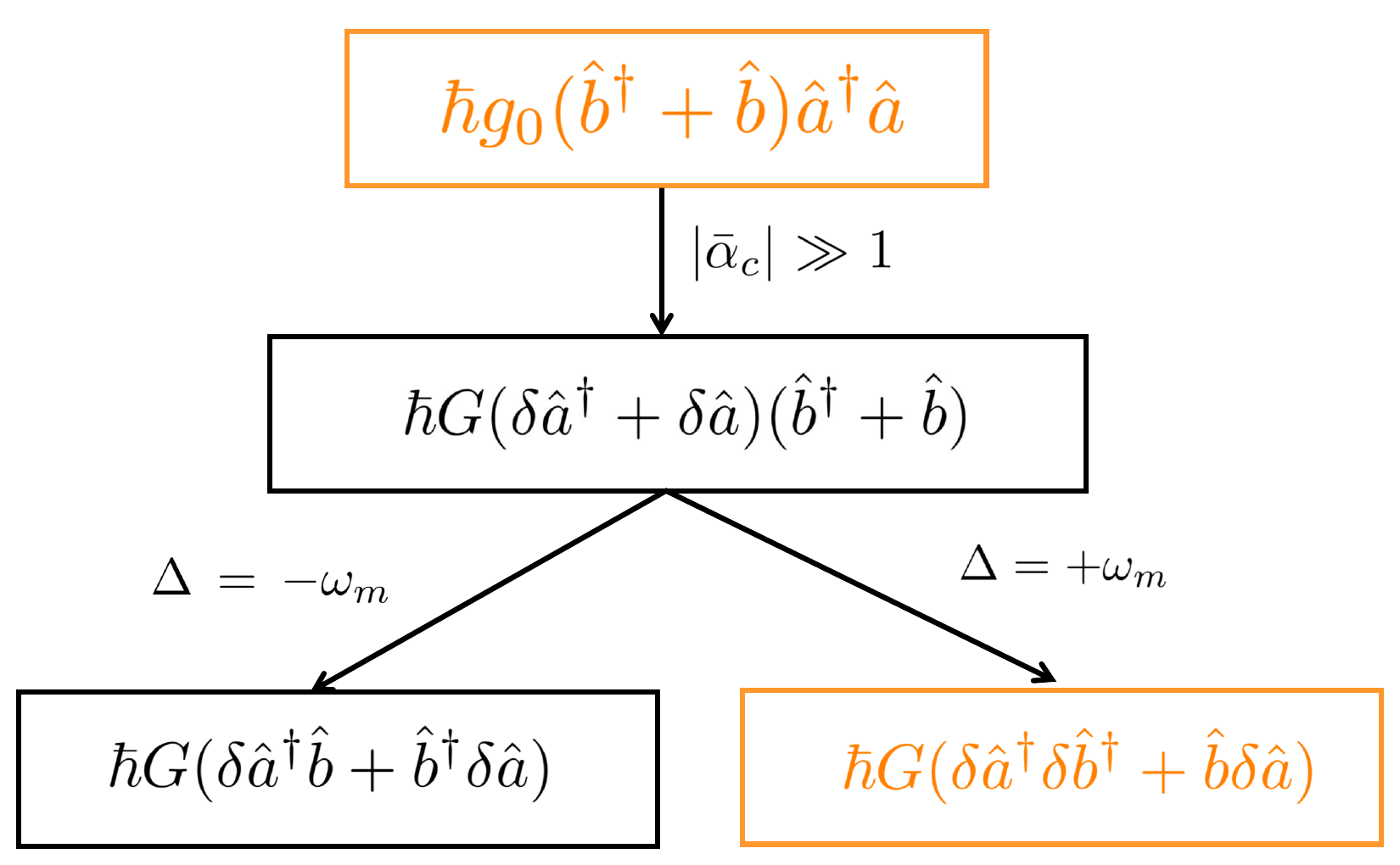}
\caption{Physical insights in the nonlinear regime of cavity-mechanical interaction 
under the strong laser limit: $|\bar{\alpha}_c|\gg 1$. When the strong laser is in 
the red-detuned regime with $\Delta=-\omega_m$, the laser controls the two-mode 
transferring process but, in the blue-detuned regime with $\Delta=+\omega_m$, the 
laser controls the exponential growth of the two modes in energies and creates the 
quantum correlation between two modes~\cite{Aspelmeyer:2014}.}
\label{fig: sche_nonlinear_under}
\end{figure}

In the nonlinear interaction regime, the physical picture of how the laser leverages 
the radiation-pressure interaction to create and stabilize the photon number and even 
the entanglement is not straightforward. However, physical insights can be gained by
examining the strong laser limit. When the amplitude of the laser is strong: 
$\bar{\alpha}_c\gg 1$, in the blue-detuned regime with $\Delta = +\omega_m$, the 
laser can modulate its frequency to create exponential growth of the energies of both
the cavity and oscillator modes, accompanied by the generation of strong quantum 
correlation between the two modes. In the red-detuned regime with $\Delta=-\omega_m$,
a switching process between the two modes occurs, which is the same as that in the 
linear interaction regime.

The blue- and red-detuned regimes have a competitive relationship with each other 
in terms of both the photon number and entanglement. In particular, in the 
blue-detuned regime, photons are excited and the rate of excitation can be larger 
than that associated with quantum dissipation to the vacuum bath. Furthermore, 
quantum entanglement is enhanced, overcoming quantum decoherence from the classical 
environment and even from the SME. However, in the red-detuned regime, no photons 
are excited and there is only a two-mode energy-transferring process that does not 
completely suppress the process of quantum dissipation to the vacuum bath, resulting 
in photon loss and eventually reducing entanglement. Stabilizing the photon number 
and entanglement requires a balance between the operations in the blue- and 
red-detuned regimes. In general, the blue-detuned regime is prone to too high photon 
levels with strong entanglement, which should be balanced by the red-detuned regime 
operation to reduce the photon number to realize our target entanglement engineering, 
as shown schematically in Fig.~\ref{fig: sche_nonlinear_under}. Overall, in the 
nonlinear interaction regime, laser driving of finite amplitude and frequency 
modulation can control the photon number and entanglement to a certain 
extent. An example is shown in Fig.~\ref{fig: WCM_nonlinear_obser_Pn}(d), where the 
RL agent finds the optimal action flow with a finite laser amplitude. Note that, the 
detuning $\Delta$ is modulated mainly in the range $\Delta \in[-2\omega_m,2\omega_m]$, 
signifying a balance between the blue- and red-detuned operations.

\subsubsection{Weak continuous measurement}

In an open quantum system, under WCM and quantum dissipation into the vacuum bath 
as well, a Wiener process occurs in the observable. More specifically, the Wiener 
process arises from the Gaussian-weighted projection over the eigenstates, which 
weakly extracts the partial information from the quantum system and induces 
stochastic disturbances in both the dynamical equation and observation. Such 
disturbances can avoid a complete quantum state collapse and provide the capability 
to extract the quantum information continuously in the time domain. However, the 
nonlinear stochastic process occurs in both quantum dynamical trajectories and the 
measurement photocurrent, making it challenging to control the quantum system 
through WCM continuously. 

For stochastic noise in the WCM photocurrent, the present cutting-edge technology 
enables the RL agent to extract quantum information through a process resembling 
noise filtering. Specifically, the observation in the reward function is the WCM 
photocurrent. We can employ $n_{\textnormal{traj}}$ quantum ensembles to reduce 
the variance and use a Gaussian filter for data pre-processing. The RL agent is 
trained to maximize the accumulated reward, which serves to average the stochastic 
term in the measurement photocurrent over time. These noise-filtering processes 
help extract information about the expected photon number and thus the target 
quantum entanglement. For the nonlinear quantum stochastic process with quantum 
dissipation, the RL agent successfully trains the laser to leverage interactions 
between the optical and mechanical modes, linear or nonlinear, to mitigate quantum 
decoherence and dissipation to some extent, as exemplified in 
Figs.~\ref{fig: measureRate_training_testing} and \ref{fig: WCM_nonlinear_obser_Pn}.

\section{Discussion} 

Exploiting machine learning for controlling quantum information systems is 
becoming a promising research realm and is attracting increasing attention.
We have developed a model-free deep-RL method for entanglement engineering. We 
demonstrated its superiority over benchmark quantum control methods in 
quantum optomechanical systems under WCM. The model-free deep-RL agent  
sequentially interacts with one or multiple parallel quantum optomechanical 
environments, collects trajectories, and updates its policy to maximize the 
accumulated reward to create and stabilize the entanglement. Both linear and 
nonlinear interacting regimes between the photons in the optical cavity and 
the phonons associated with the mechanical oscillator in the cavity have been 
studied. In particular, for linear interactions, the PPO agent directly observes 
the WCM photocurrent and delivers better performance compared with the benchmark 
Bayesian and random control methods in the framework of measurement-based feedback 
control. The performance of deep RL control is tolerant to randomness when initially 
the system is in some mixed state. For nonlinear interactions, both the model-free 
PPO and recurrent PPO agents have been tested, where the first was utilized to 
generate the time series of the target of the expected photon number, and the 
second one was employed to control entanglement according to an objective. Because 
of the high degree of randomness in the SME originating from ten stochastic collapse 
operators, only the observation of the expected photon number is feasible in the 
nonlinear interaction regime.

More specifically, linear interactions can naturally limit the excitation in the 
energy levels, providing a mechanism to directly create the entangled Bell states 
under the premise of strong laser approximation in the red-detuned regime. A 
disadvantage is that its performance is sensitive to the coupling of the vacuum 
or thermal bath, even when the decay rate is small (e.g., $\kappa=0.01\,\omega_m$ 
with $G\in[-5,5]\,\omega_m$). This phenomenon is in fact quite common in quantum
systems. For instance, in systems with magnon-photon coupling~\cite{Yuan:2020}, 
steady Bell states can ideally be generated in the PT-broken phase without 
dissipation while the entanglement is reduced when the decay rate is 
not negligible. Another issue with linear interactions is that the time scale 
associated with generating entangled Bell states~\cite{Li:2023} tends to be much 
shorter than the inverse of the coupling strength about the higher-order exceptional 
points in a system of coupled non-Hermitian qubits with energy loss while the 
maximum entanglement can only last for a short instant.

In contrast, nonlinear interactions can create and stabilize entanglement and 
are more robust to the disturbance from the vacuum bath even with a relatively
large decay rate, e.g., $\kappa=0.1\,\omega_m$  with the strong coupling 
$g_0=0.2\,\omega_m$, so $g_0/\kappa=2>1$ to stipulate the nonlinear 
effect~\cite{Aspelmeyer:2014}. Potentially, systems with nonlinear coupling thus 
can outperform those with linear interactions. A caveat is that, in nonlinear 
optomechanical systems, there is limited experimentally accessible observation. 
In fact, the relationship between experimental observables and entanglement in 
nonlinear quantum optomechanical systems has not been well understood, rendering 
challenging to choose a feasible observable to control entanglement. We have 
partially relied on the numerical method to create and stabilize entanglement, 
based on the numerical relation between entanglement and the expected photon number 
discovered by the deep RL. Another difficulty is that the nonlinear interaction 
can readily excite the system to high quantum states, which we have overcome by 
designing a proper reward function.

A previous work~\cite{hill:2008} studied the acceleration of entanglement generation 
through feedback weak measurement for two qubits in a four-dimensional Hilbert space,
where coupling to a vacuum or thermal bath was not taken into account, nor the 
interactions between the two qubits, and the control protocol required prior 
knowledge about the system such as the decoherence-free subspace. In addition, 
complete observation was needed to design the local Hamiltonian feedback to speed 
up entanglement. This is in fact a model-based approach. In another 
study~\cite{diotallevi:2024}, steady-state entanglement between two qubits was
achieved using a continuous feedback control method, where the feedback protocol 
design was informed by a detailed model of the system’s dynamics. In contrast, our 
work creates and stabilizes a two-mode entangled state about a predetermined level 
of entanglement for both linear and nonlinear interactions via model-free 
reinforcement learning, with the respective dimensions of the Hilbert space being 
four and one hundred.

Gate-based control uses quantum logic gates to manipulate qubit states and generate entanglement, typically relying on Von Neumann projective measurement, which requires $10^{3} - 10^{6}$ repetitions to estimate an observable. In contrast, weak continuous measurement employs Gaussian projections, eliminating repeated system preparation and requiring only a single preparation for the entire time horizon. While adiabatic passage achieves high-fidelity states through slow evolution under a time-dependent Hamiltonian, its long duration remains a limitation. Advances in counterdiabatic driving~\cite{yao:2021} aim to address this limitation but remain model-based, relying on predefined mathematical models. Similarly, dissipation engineering~\cite{harrington:2022}, which leverages tailored dissipation to stabilize desired states, is also inherently model-based. The RL, by contrast, offers a model-free approach, requiring no prior knowledge of mathematical models and needing only a single initial preparation for each episode. Weak continuous measurement further reduces repetitions to one per time step, significantly lowering quantum resource use and running time, making the model-free framework efficient and adaptable across quantum models.
  
Our work suggests the possibility of exploiting multi-agent RL through parallel 
computation to stabilize entanglement. The agents leverage the decentralized 
structure of the task and share information via communication. Saliently, if 
several agents fail in a multiagent system, the remaining agents can take over 
some of their tasks. In principle, our control framework can be extended to 
multi-agent RL for multi-mode entanglement engineering of a quantum black box.

In our work, all methods discussed operate within the framework of Markovian feedback, particularly RL based on Markov Decision Processes (MDPs). This means that the action at the next time step is determined only by the state at the current time step.  However, the time delay is a fundamental challenge in any feedback control system, including quantum systems, arising from the inevitable lag between measurement, processing, and feedback application~\cite{deng:2022,chembo:2019,fridman:2022}. This delay often stems from the time required to extract information from the quantum system, such as through photodetectors or other measurement apparatus. To address this issue, incorporating delay-compensation techniques into the feedback controller design is essential to account for and mitigate the effects of known delays.

\section{Methods}\label{sec:methods}

\subsection{Stochastic Master Equation}

An experimental optomechanical system is effectively an open quantum system 
interacting with the vacuum bath under WCM with the 
operators~\cite{essig:2021,porotti:2022} $\hat{C}_n\equiv \sqrt{\eta}\hat{P}_n$,
where $\hat{P}_n=|n\rangle\langle n|$ with $n=0,1$ (linear) or $n=0,1,\ldots,9$ 
(nonlinear) is the measurement operator on the Fock state and $\eta$ denotes the 
measurement rate. The quantum dynamics of this system are described by the 
stochastic master equation 
(SME)~\cite{blaquiere:1987,porotti:2022,bowen:2015,wiseman:2009,jacobs:2006}:
\begin{align} \label{eq:sme}
    d\rho &= \frac{1}{i\hbar}[\tilde{H},\rho]dt+ \mathcal{L}_{env}\;\rho dt\nonumber\\
    &+\sum_n \mathcal{D}(\hat{C}_n)\rho dt+ \sum_n \mathcal{H}(\hat{C}_n)\rho dW_n,
\end{align}
where the Hamiltonian is $\tilde{H}=\tilde{H}_{bs}$ or $\tilde{H}_{nl}$ and $\rho$ is 
a density operator in the Hilbert space. Under the Born-Markov 
approximation~\cite{nathan:2020,bai:2021}, which requires the system-bath coupling to 
be weak and the correlation time of the bath to be much shorter than a characteristic 
timescale of system-bath interactions, the Markovian master equation, i.e., the first 
two terms in the right-hand side of Eq.~\eqref{eq:sme}, has the Lindblad 
form~\cite{nathan:2020}. At absolute zero temperature, the following environmental 
operator $\mathcal{L}_{env}\;\rho$ can be introduced to describe the coupling 
between the system and vacuum bath: 
$\mathcal{L}_{env}\;\rho = \kappa \mathcal{D}(\hat{a})\rho + \gamma \mathcal{D}(\hat{b})\rho$, 
where the cavity and oscillator modes are coupled to the vacuum bath with the 
strength $\kappa$ and $\gamma$, respectively~\cite{qian:2012}. The deep RL results 
in the Lindblad master equation with the nonlinear interaction are presented in 
Appendix~\ref{sec:appendix_RL_nonlinear}.

The WCM process described by the last two terms in the right-hand side of 
Eq.~\eqref{eq:sme} is nonlinear and Markovian in the unconditional master 
equation~\cite{jacobs:2006} in $\rho$. Under WCM, a Wiener process $dW$ with a 
Gaussian distribution~\cite{jacobs:2006} arises from the Gaussian-weighted projection 
over the eigenstates that allows the quantum information to be extracted continuously 
in the time domain, subject to stochastic disturbances in the last term of 
Eq.~\eqref{eq:sme} and quantum decoherence in the penultimate term of 
Eq.~\eqref{eq:sme}. (Appendix~\ref{sec:appendix_sme} provides a detailed derivation 
of the SME.) The Lindblad operator $\mathcal{D}$ and the measurement superoperator 
$\mathcal{H}$ in Eq.~\eqref{eq:sme} are given by 
\begin{align} \nonumber
	\mathcal{D}(\hat{A})\rho &\equiv\hat{A}\rho\hat{A}^\dagger-\frac{1}{2}(\hat{A}^\dagger\hat{A}\rho+\rho\hat{A}^\dagger\hat{A}), \\ \nonumber
	\mathcal{H}(\hat{A})\rho &\equiv\hat{A}\rho+\rho\hat{A}^\dagger-\langle\hat{A}+\hat{A}^\dagger\rangle\rho, 
\end{align}
with $\langle\hat{A}\rangle\equiv\textnormal{Tr}[\hat{A}\rho]$. The two operators 
serve to weakly drive the quantum state into the corresponding eigenstates to some 
degree. 

\subsection{Implementation details of deep RL}\label{subsec:details_RL}

For simulating the linear or nonlinear quantum optomechanical system described by 
Eq.~\eqref{eq:sme}, we use the ``taylor1.5'' solver from the SME solver in the 
QuTip's package~\cite{johansson:2012} with the tolerance $\textnormal{tol}=10^{-6}$ 
and time step size $dt=0.01\,\omega_{m}^{-1}$. The measurement current is simulated 
with the ``homodyne'' method, and the custom environment is constructed by the 
open-source platform OpenAI-Gym~\cite{brockman:2016}. For RL simulations, we 
construct the PPO agent~\cite{schulman:2017} by 
``stable-baselines3''~\cite{stable-baselines3} in the A2C~\cite{mnih:2016} settings, 
where stochastic policy (actor) and the value function (critic) are modeled by two 
independent neural network function approximators, i.e., a set of fully connected 
feed-forward networks of dimensions $256\times 128\times 64$ and the hyperbolic 
tangent nonlinear activation function for each hidden layer. For the nonlinear 
quantum optomechanical configuration, in the target-utilization phase, the recurrent 
PPO agent outperforms the PPO agent, where both independent critic and actor networks 
are MLPs followed by one independent layer of LSTM with $256\times 128\times 64$ 
fully connected networks and $256$ hidden states. More details, especially in the table of Gaussian kernel, are described in Appendix~\ref{sec:appendix_DRL}.

\section*{Acknowledgments}

We thank Dr.~Kanu Sinha for discussions and comments. This work was supported by 
the Air Force Office of Scientific Research under Grant No.~FA9550-21-1-0438 and 
by the Office of Naval Research under Grant No.~N00014-24-1-2548. The Quantum 
Collaborative, led by Arizona State University, also provided valuable expertise 
and resources for this work through seed funding. The Quantum Collaborative connects 
top scientific programs, initiatives, and facilities with prominent industry partners 
to advance the science and engineering of quantum information science.

\section*{AUTHOR DECLARATIONS}
\subsection*{Conflict of Interest}
\vspace*{-0.1in}
The authors have no conflicts to disclose.

\section*{Author Contributions}
\vspace*{-0.1in}
L.-L. Y., C. A., J. L., and Y.-C. L. designed the research project, the models, and
methods. L.-L. Y. performed the computations. L.-L. Y., C. A., J. L., and Y.-C. L.
analyzed the data. L.-L. Y. and Y.-C. L. wrote and edited the manuscript.

\section*{Data availability}

The codes used in this paper can be found in the
repository: https://github.com/liliyequantum/Entanglement-engineering-by-RL~\cite{our_codes}. The data generated in this study about training results can be found in this Zenodo repository: https://doi.org/10.5281/zenodo.12584159~\cite{our_data}. 

\appendix

\section{Background related to our work}\label{sec:appendix_background_work}
\paragraph*{Quantum control.}
Quantum control~\cite{wiseman:2009} is essential to quantum engineering and 
technology~\cite{rosencher:1996,iannaccone:2018,bohn:2017}, where open-loop 
control~\cite{paz:2014} has been successfully demonstrated with methods such as 
gradient-ascent pulse engineering (GRAPE)~\cite{Machnes:2011} in spin 
systems~\cite{dolde:2014}, coupled qubits~\cite{Sporl:2007}, Jaynes-Cummings 
systems~\cite{heeres:2017}, and qubit-cavity lattices~\cite{Fisher:2010}. Recently, 
the open-loop GRAPE algorithm has been extended to feedback GRAPE~\cite{porotti:2023}
based on gradient ascent of quantum dynamics for state engineering under strongly 
stochastic measurement. Open-loop control, however, requires a differentiable model 
of the quantum dynamics that may not always be available. In realistic situations 
where such a model is not available, closed-loop feedback control strategies 
conditioned on experimental measurement outcomes can be applied. Combined with 
data-driven machine learning, feedback control has been implemented in experiments 
in a model-free fashion~\cite{mnih:2015,arulkumaran:2017,sutton:2018}.

\paragraph*{Deep reinforcement learning.}
In general, RL is a machine-learning paradigm based on a trial-and-error learning 
process, incorporating traditional optimal control to maximize the accumulated 
reward. The use of deep neural networks in the learning process leads to deep RL, 
which explores and exploits the available measurement data to search for a 
globally optimal policy. In deep RL, many algorithms are available such as 
deep-Q network (DQN)~\cite{mnih:2015}, deep deterministic policy 
gradient (DDPG)~\cite{lillicrap:2015}, and trust region proximal optimization 
(TRPO)~\cite{wu:2017}. A state-of-the-art deep RL algorithm for continuous control 
is proximal policy optimization (PPO)~\cite{schulman:2017}, whose performance can 
exceed that of TRPO. Incorporating recurrent neural networks~\cite{hochreiter:1997} 
into the PPO algorithm leads to improved performance~\cite{pleines:2022}. In recent 
years, measurement-based feedback control with deep RL has been applied to quantum 
systems for tasks such as quantum error correction for discrete 
gates~\cite{fosel:2018}, state preparation and stabilization for a single 
particle~\cite{wang:2020,Borah:2021,sivak:2022,porotti:2022} with an unstable 
potential~\cite{wang:2020} or a double-well potential~\cite{Borah:2021}, 
discrimination between entangled states~\cite{cao:2023} for quantum meteorology, 
and long-distance entanglement distribution on quantum networks~\cite{haldar:2024}. 
Experimentally, time scales of the RL action sequences shorter than the coherence 
time of the underlying quantum system have been realized, rendering feasible 
real-time deep-RL feedback control~\cite{reuer:2022}.

\paragraph*{Quantum measurement.} In quantum systems, projective measurement can 
be used to extract the full information about the quantum state but, as a back 
action, the quantum state will collapse after the measurement~\cite{Riste:2012}. To 
avoid a complete collapse, one can exploit weak 
measurements~\cite{Lundeen:2012,Smith:2004}, in which the probe is weakly coupled 
to the system to yield partial information about the quantum state. Examples of 
weak measurements include continuous monitoring~\cite{Yang:2023} of 
driven dissipative quantum-optical systems - a basic component of quantum 
meteorology~\cite{Giovannetti:2006,giovannetti:2011}. A form of weak measurement, 
the so-called weak continuous measurement (WCM), is fundamental to a broad range 
of applications. For example, WCM has been used to detect the quadrature 
operators~\cite{puentes:2009}, Wigner~\cite{puentes:2009} and Husimi Q 
functions~\cite{kanem:2005} with a homodyne apparatus~\cite{lvovsky:2009}, 
rendering observing both pure~\cite{lundeen:2011} and mixed~\cite{Lundeen:2012} 
quantum states experimentally feasible. WCM has been experimentally implemented 
by a weak-field homodyne detector~\cite{Thekkadath:2020,puentes:2009,lvovsky:2009} 
to measure the photon-number statistical distribution over the Fock basis. 
In another example, WCM has been realized in an atomic spin 
ensemble~\cite{Smith:2004} via Faraday rotation of an off-resonance probe beam to 
create and probe nonclassical spin state and dynamics. The concept of WCM has also 
been used to develop fundamental theories, such as Heisenberg's 
measurement-disturbance relationship~\cite{Rozema:2012} and error-disturbance 
uncertainty relation~\cite{Kaneda:2014}. Because of the typical time scales of 
the quantum dynamics, WCM cannot be regarded as occurring 
instantaneously~\cite{jacobs:2006}. Theoretically, the impact of WCM on the 
underlying quantum system can be described by the stochastic master 
equation~\cite{jacobs:2006}.

\section{Quantum optomechanical system}\label{sec:appendix_Hamiltonian}

The standard Hamiltonian of a quantum optomechanical system in the rotating frame 
of the laser is given by~\cite{liu:2013,qian:2012}
\begin{align}
    \tilde{H} &= -\hbar\Delta \hat{a}^{\dagger}\hat{a} + \hbar\omega_{m}\hat{b}^{\dagger}\hat{b}+\hbar g_0(\hat{b}^{\dagger}+\hat{b})\hat{a}^{\dagger}\hat{a}\nonumber\\
    &~~~~~~~~~~~+\hbar(\alpha_{L}\hat{a}^{\dagger}+\alpha_{L}^{*}\hat{a}),
\end{align}
where $\hat{a},\hat{b}$ $(\hat{a}^{\dagger},\hat{b}^{\dagger})$ are the annihilation 
and creation operators of the optical cavity and mechanical mode, respectively. The 
frequency detuning is $\Delta \equiv \omega_L - \omega_c$, where $\omega_L$ is the 
frequency of the driven laser and $\omega_c$ is the intrinsic frequency of the 
cavity. The nonlinear coupling $g_0$ between the single cavity and mechanical mode 
arises from the frequency dispersion relationship with respect to the displacement 
$\hat{q}$ of the mechanical mode. The complex amplitude of the driven electromagnetic 
field is denoted as $\alpha_L$. A detailed description of how the Hamiltonian is 
derived is as follows.

Consider a single optical cavity and a mechanical mode (with a movable mirror). 
The resonant frequency of the cavity mode is controlled by the displacement of the
movable end-mirror $\omega_c(\hat{q})$ or the length of the cavity, which can be 
expanded to the first order about the intrinsic frequency $\omega_c(\hat{q}=0)$ 
of the cavity, leading to the following nonlinear coupling term:
\begin{align}
        \hat{H}_0 &= \hbar\omega_c(\hat{q}) \hat{a}^{\dagger}\hat{a}+ \hbar\omega_{m}\hat{b}^{\dagger}\hat{b}\nonumber\\
        & = \hbar(\omega_c+(\partial\omega_c(q)/\partial q)\hat{q})\hat{a}^{\dagger}\hat{a}+\hbar\omega_{m}\hat{b}^{\dagger}\hat{b}\nonumber\\
        & = \hbar\omega_c\hat{a}^{\dagger}\hat{a}+\hbar\omega_{m}\hat{b}^{\dagger}\hat{b}+\hbar g_0 \hat{a}^{\dagger}\hat{a}(\hat{b}^\dagger+\hat{b}),
\end{align}
where $g_0 \equiv (\partial\omega_c(q)/\partial q)q_{\textnormal{zpf}}$ is the 
single-photon optomechanical coupling strength and the position operator of the 
mechanical mode is $\hat{q}\equiv (\hat{b}+\hat{b}^\dagger)q_{\textnormal{zpf}}$ 
with $q_{\textnormal{zpf}}=\sqrt{\hbar/(2m\omega_{m})}$ being the mechanical 
zero-point fluctuations. The radiation pressure force is acted on the mechanical 
resonator by the photon number operator multiplying the displacement operator 
$\hat{q}$.

The Hamiltonian $\hat{H}=\hat{H}_0+\hat{H}_{driven}$ in the rotating frame is 
defined as~\cite{bowen:2015}:
\begin{align}
\tilde{H} = \hat{U}^\dagger\hat{H}\hat{U}-\hat{A}
\end{align}
with $\hat{U}\equiv \exp{(-i\omega_L\hat{a}^\dagger\hat{a}\;t)}$ and 
$\hat{A}\equiv\hbar\omega_L\hat{a}^\dagger\hat{a}$. Using the following identities:
\begin{align}
        &\exp{(i\omega_L\hat{a}^\dagger\hat{a}\;t)}\hat{a}\exp{(-i\omega_L\hat{a}^\dagger\hat{a}\;t)}=\hat{a}\exp{(-i\omega_L t)},\nonumber\\
        &\exp{(i\omega_L\hat{a}^\dagger\hat{a}\;t)}\hat{a}^\dagger\exp{(-i\omega_L\hat{a}^\dagger\hat{a}\;t)}=\hat{a}^\dagger\exp{(i\omega_L t)},
\end{align}
we have 
\begin{align} \nonumber
\hat{U}^\dagger\hat{a}^\dagger\hat{a}\hat{U} = \hat{a}^\dagger\hat{a}. 
\end{align}
In the rotating frame, with the detuning $\Delta\equiv\omega_L-\omega_c$, we then have
\begin{align}
\tilde{H}_0 = -\hbar\Delta\hat{a}^{\dagger}\hat{a}+\hbar\omega_{m}\hat{b}^{\dagger}\hat{b}+\hbar g_0 \hat{a}^{\dagger}\hat{a}(\hat{b}^\dagger+\hat{b}).
\end{align}
The quantized electromagnetic field can be written as
\begin{equation}
        \hat{H}_{driven} =  \hbar\left[\alpha_L \exp(-i\omega_L t)\hat{a}^\dagger+\alpha_L^{*}\exp(i\omega_L t)\hat{a}\right].
\end{equation}
Through the unitary transformation, we obtain
\begin{align}
\hat{U}^\dagger\hat{H}_{driven}\hat{U}=\hbar\alpha_L\hat{a}^\dagger+\hbar\alpha_L^{*}\hat{a}.
\end{align}
Finally, the total Hamiltonian driven by the electromagnetic field in the rotating 
frame is given by 
\begin{align}
        \tilde{H} &= \hat{U}^\dagger(\hat{H}_0+\hat{H}_{driven})\hat{U}-\hat{A}\nonumber\\
        &=\tilde{H}_0 + \hat{U}^\dagger\hat{H}_{driven}\hat{U}\nonumber\\
        &= -\hbar\Delta\hat{a}^{\dagger}\hat{a}+\hbar\omega_{m}\hat{b}^{\dagger}\hat{b}+\hbar g_0 \hat{a}^{\dagger}\hat{a}(\hat{b}^\dagger+\hat{b})\nonumber\\
        &~~~~~~~~~~~~~+\hbar(\alpha_L\hat{a}^\dagger+\alpha_L^{*}\hat{a}).
\end{align}

\section{Quantum stochastic master equation} \label{sec:appendix_sme}

The starting point is von Neumann equation, which governs the unitary evolution of 
the density matrix and is given by 
\begin{equation} \label{eq:Von_Neumann}
    \Dot{\rho} = \frac{1}{i\hbar}[\hat{H},\rho]\equiv \mathcal{L}\rho,
\end{equation}
where $\mathcal{L}$ is the Liouvillian superoperator. Equation~\eqref{eq:Von_Neumann}
can be derived from the Schr\"{o}dinger equation and its conjugate: 
\begin{align}
        i\hbar \frac{\partial}{\partial t}|\psi\rangle &= \hat{H}|\psi\rangle,\nonumber\\
        -i\hbar \frac{\partial}{\partial t}\langle\psi|&= \langle\psi|\hat{H},
\end{align}
with Hermitian Hamiltonian $\hat{H}^\dagger=\hat{H}$. Since the density matrix is 
defined as a mixture of quantum states,
$\rho=\sum_{i}P_i|\psi_i\rangle\langle\psi_i|$ with $\sum_i P_i = 1$, we have
\begin{align}
        i\hbar\Dot{\rho}&=\sum_i P_i (i\hbar |\Dot{\psi_i}\rangle)\langle\psi_i|-\sum_i P_i |\psi_i\rangle(-i\hbar\langle\Dot{\psi}_i|)\nonumber\\
        & = \sum_i P_i \hat{H}|\psi_i\rangle\langle\psi_i|-\sum_i P_i|\psi_i\rangle\langle\psi_i|\hat{H}\nonumber\\
        &=\hat{H}\rho-\rho\hat{H} = [\hat{H},\rho],  
\end{align}
where $\partial\rho/\partial t\equiv \Dot{\rho}$ and 
$\partial |\psi\rangle/\partial t\equiv\Dot{|\psi\rangle}$.

The dynamics of a quantum system interacting with the vacuum bath under the continuous
measurement of the observable $\hat{c}$ are described by the general stochastic master equation(SME)~\cite{bowen:2015,wiseman:2009,jacobs:2006}:
\begin{align}
    d\rho &= \frac{1}{i\hbar}[\hat{H},\rho]dt+ \mathcal{L}_{env}\;\rho dt+\mathcal{D}(\hat{c})\rho dt + \mathcal{H}(\hat{c})\rho dW,
\end{align}
where $\mathcal{L}_{env}\;\rho$ is the interaction between the system and vacuum 
bath, which is given by 
\begin{align} \label{eq:L_env}
\mathcal{L}_{env}\;\rho = \kappa \mathcal{D}(\hat{a})\rho + \gamma \mathcal{D}(\hat{b})\rho, 
\end{align}
and $dW$ corresponds to a Wiener process with a Gaussian distribution. Concretely, 
both the cavity and the oscillator modes are coupled to the vacuum bath with the 
coupling strengths $\kappa$ and $\gamma$, respectively, where the bath is at the 
absolute zero temperature. In Eq.~\eqref{eq:L_env}, the symbols $\mathcal{D}$ and 
$\mathcal{H}$ denote the Lindblad and measurement superoperators, respectively, 
which are given by
\begin{align}
\mathcal{D}(\hat{c})\rho&\equiv\hat{c}\rho\hat{c}^\dagger-\frac{1}{2}(\hat{c}^\dagger\hat{c}\rho+\rho\hat{c}^\dagger\hat{c}), \\
\mathcal{H}(\hat{c})\rho&\equiv\hat{c}\rho+\rho\hat{c}^\dagger-\langle\hat{c}+\hat{c}^\dagger\rangle\rho.
\end{align}
The actions described by the two superoperators can drive the quantum state into an 
eigenstate of the observable $\hat{c}$ to some degree. Pertinent to this process is 
WCM~\cite{jacobs:2006}. To understand WCM, we begin with the von Neumann measurement.
    
The set of eigenstates of an observable forms an orthonormal basis in the Hilbert 
space: $\{|n\rangle:n=1,...,n_{\textnormal{max}}\}$. Any pure quantum state can be 
completely expanded as $|\psi\rangle=\sum_n c_n|n\rangle$ with the probability 
distribution $|c_n|^2$ over the basis $\{|n\rangle\}$. The von Neuman measurement, 
after which the quantum state will be completely projected onto one of the 
eigenstates of the observable, gives complete information about the collapsed quantum 
state. More specifically, the measurement can be described by a set of projection 
operators $\{P_n=|n\rangle\langle n|\}$ based on the orthonormal basis of the 
observable. If the initial state is $\rho = |\psi\rangle\langle\psi|$, the 
probability of obtaining the $n$th eigenvalue will be $\textnormal{Tr}[P_n\rho P_n]$ 
with the final state given by
\begin{equation}
        \rho_f=\frac{P_n\rho P_n}{\textrm{Tr}[P_n\rho P_n]}=|n\rangle\langle n|.
\end{equation}
While von Neumann measurement provides complete information for the collapsed 
quantum state after being measured since the state has collapsed to an eigenstate 
of the observable after the projective measurement, it is not the only kind of 
measurement. Other methods can reduce the uncertainty of the observable but often
fail to remove all of it. Such measurements can extract only partial information 
about the quantum system. 
    
In principle, we can choose a set of $m_{\textnormal{max}}$ operators $\Omega_m$ 
with the restriction 
\begin{align} \nonumber
\sum_{m=1}^{m_{\textnormal{max}}}\Omega^\dagger_m\Omega_m = I, 
\end{align}
where the number $m_{\textnormal{max}}$ of elements can be larger than the 
dimension of the Hilbert space which they act in. A measurement with $N$ possible 
outcomes can be designed for 
\begin{equation}
        \rho_f=\frac{\Omega_m\rho \Omega^\dagger_m}{\textrm{Tr}[\Omega_m\rho \Omega^\dagger_m]},
\end{equation}
with the probability $\textnormal{Tr}[\Omega_m\rho \Omega^\dagger_m]$. For example, 
the probability of the observation in the range $[a,b]$ is given by
\begin{align}     
P(m\in[a,b])=\sum_{m=a}^{b}\textrm{Tr}[\Omega_m\rho \Omega^\dagger_m]= \textrm{Tr}[\sum_{m=a}^{b}\Omega_m\rho \Omega^\dagger_m].
\end{align}
The measurement, associated with a positive operator 
$M=\sum_{m=a}^{b}\Omega^\dagger_m\Omega_m$ with every subset in the range 
$m\in[1,m_{\textnormal{max}}]$, is called a positive operator-valued measure (POVM).
   
POVMs can describe weak measurements, where only partial information is 
extracted from the measurement by the Gaussian weighted sum over all eigenstates 
of the observable:
\begin{equation}
       \Omega_m = \frac{1}{\mathcal{N}}\sum_n e^{-k(n-m)^2/4}|n\rangle\langle n|,
\end{equation}
with the normalization constant $\mathcal{N}$ that satisfies the constraint 
$\sum_{m=-\infty}^{\infty}\Omega^\dagger_m\Omega_m=I$. Suppose no information is 
obtained before the measurement and the initial state is completely mixed as 
$\rho\propto I$, then the observation is a random variable with Gaussian 
distribution. After the measurement, the state becomes
\begin{align}
       \rho_f&=\frac{\Omega_m\rho \Omega^\dagger_m}{\textrm{Tr}[\Omega_m\rho \Omega^\dagger_m]} = \frac{1}{\mathcal{N}}\sum_n e^{-k(n-m)^2/2}|n\rangle\langle n|.
\end{align}
This indicates that, when the initial state $\rho$ is an equal probability 
distribution over all eigenstates, the state after the weak measurement has a
Gaussian distribution over all the eigenstates, where the mean value of the Gaussian 
weights corresponds to an eigenstate and the distribution spreads with a finite 
uncertainty. Consequently, only partial information can be extracted from this kind 
of measurement, because it only partially projects onto an eigenstate of the 
observable with uncertainty. The standard deviation of the final state is 
$1/\sqrt{k}$. The larger the measurement strength $k$, the more complete 
information can be extracted with reduced uncertainty about the quantum state, 
leading to strong measurement. On the contrary, a small measurement strength 
generates weak measurement. 

We can now describe WCM. In general, continuous measurement means that information 
is continually extracted from a system over time. To realize WCM, time is divided 
into a series of intervals of size $\Delta t$, and a weak measurement is carried out
in each interval. The Hermitian observable is denoted as $\hat{O}$, and the 
measurement operator with the index $\alpha$ is given by
\begin{align}\label{eq:gaussian_projector}
       \hat{A}(\alpha) = \left(\frac{4k\Delta t}{\pi}\right)^{1/4}\int_{-\infty}^{\infty} e^{-2k\Delta t(O-\alpha)^2}|O\rangle\langle O| dO,
\end{align}
where the measurement strength is determined by $k$ and $\Delta t$. If we set 
$\Delta t=dt$, then it is a WCM. The mean of the continuous index $\alpha$ is
\begin{align}
       \langle\alpha\rangle=\int_{-\infty}^{\infty}\alpha Tr[\hat{A}^\dagger(\alpha)\hat{A}(\alpha)|\psi\rangle\langle\psi|]d\alpha=\langle \hat{O}\rangle.
\end{align}
The probability distribution of $\alpha$ is
\begin{align} \label{eq:P_alpha}
       P(\alpha) &= Tr[\hat{A}^\dagger(\alpha)\hat{A}(\alpha)|\psi\rangle\langle\psi|]\nonumber\\
       & = \sqrt{\frac{4k\Delta t}{\pi}}\int_{-\infty}^{\infty}|\psi(O)|^2 e^{-4k\Delta t(O-\alpha)^2}dO.
\end{align}
The value of $\Delta t$ is infinitesimal due to the inherent property of the WCM. 
As a result, the exponential term in Eq.~\eqref{eq:P_alpha} is a slow oscillation 
compared with the wave function under the variable $O$. Based on this, the wave 
function can be approximated as $|\psi(O)|^2\approx\delta(O-\langle O\rangle)$ and 
we have
\begin{align}
       P(\alpha)\approx \sqrt{\frac{4k\Delta t}{\pi}} e^{-4k\Delta t(\alpha-\langle O\rangle)^2}.
\end{align}
Effectively, $\alpha$ is a stochastic quantity:
\begin{align}\label{eq:alpha_s}
       \alpha_s = \langle \hat{O}\rangle + \frac{\Delta W}{\sqrt{8k}\Delta t},
\end{align}
where $\Delta W$ is a zero-mean, Gaussian random variable with variance $\Delta t$. 
The time evolution of the quantum state under WCM is given by
\begin{align}
       |\psi(t+\Delta t)\rangle&\propto\hat{A}(\alpha)|\psi(t)\rangle\propto e^{-2k\Delta t(\alpha-\hat{O})^2}|\psi(t)\rangle.
\end{align}
Substituting Eq.~(\ref{eq:alpha_s}) into this equation, applying Taylor's expansion 
into the exponential term to first order in $\Delta t$ and defining 
$|\psi(t+dt)\rangle\equiv|\psi(t)\rangle+d|\psi\rangle$, we obtain the following
stochastic differential equation:
\begin{align}
       d|\psi\rangle=\{-k(\hat{O}-\langle\hat{O}\rangle)^2dt+\sqrt{2k}(\hat{O}-\langle\hat{O}\rangle)dW\}|\psi(t)\rangle.
\end{align}
Defining $\rho(t+dt)\equiv\rho(t)+d\rho$, we have
\begin{align}
       d\rho &= (d|\psi\rangle)\langle\psi|+|\psi\rangle(d\langle\psi|)+(d|\psi\rangle)(d\langle\psi|)\nonumber\\
       &=-k[\hat{O},[\hat{O},\rho]]dt + \sqrt{2k}(\hat{O}\rho+\rho\hat{O}-2\langle\hat{O}\rangle\rho)dW.
\end{align}
If we redefine the observable as 
\begin{align} \nonumber
	\hat{c}\equiv\sqrt{\eta}\hat{O}\equiv\sqrt{2k}\hat{O}, 
\end{align}
the first term can be rewritten as
\begin{align}
       [\hat{c}\rho\hat{c}-\frac{1}{2}(\hat{c}^2\rho+\rho\hat{c}^2)]dt
\end{align}
and the second term is
\begin{align}
      (\hat{c}\rho+\rho\hat{c}-2\langle\hat{c}\rangle\rho) dW,
\end{align}
which are consistent with the Lindblad operator $\mathcal{D}$ and the measurement superoperator 
$\mathcal{H}$ in the SME from the Method part in the main text, respectively. Here, the measurement rate $\eta$ is 
proportional to the measurement strength $k$.

The measurement rate is proportional to the measurement strength. The measurement strength quantifies the extent to which a measurement process interacts with a quantum system and perturbs its state. Measurement efficiency quantifies the fraction of information extracted from the quantum system during the measurement that is successfully captured and used to update the quantum state. In general, the measurement efficiency is less than 100\% due to factors such as detector inefficiencies and signal losses. For convenience, we incorporate the measurement efficiency into the measurement rate to define an effective measurement rate $\eta$, which accounts for both the intrinsic measurement strength and the efficiency of the process~\cite{wiseman:2009}. Additionally, the gain in the measurement current can be adjusted to equivalently mitigate the effects of measurement inefficiency, effectively restoring the information lost due to imperfect efficiency~\cite{wang:2020,Borah:2021}.

\section{Reinforcement learning (RL) in linear quantum optomechanics}\label{sec:appendix_RL_linear}

Based on the demonstration in the main text about RL in linear quantum optomechanics. This section gives the corresponding details about reinforcement learning for the linear system. During online training, given a fixed training episode length, e.g., 
${\rm Episode} = 3000$, the RL agent bootstraps itself by executing the procedure 
described in Appendix~\ref{subsec:appendix_PPO}. In the 
initial preparation process, $\mathbb{N}$ identical and independent quantum 
optomechanical environments ($\mathbb{N}$ parallel environments) are prepared, 
where $\mathbb{N}=5$. In addition, the agent, which has two independent neural 
networks: actor and critic, is also initialized. The initial quantum state is $|\psi\rangle=|10\rangle$ or  $\rho=(1-p)|10\rangle\langle 10|+ p|01\rangle\langle 01|$ with $p\in[0,1]$ and the quantum environments are governed by the SME.

In episodic learning, the quantum environments are reset after each episode. For 
each set of $\mathbb{Z}$ episodes (e.g., $\mathbb{Z}=5$), the agent obtains the 
observation $O_t$ about the photon number and the reward value $R_t=-|O_t-0.5|$ from $\mathbb{N}$ quantum environments, 
and independently acts on them by the current stochastic policy 
$\pi(G_t|O_t;\pmb{\theta})$. Essentially, the policy is the conditional 
probability distribution on the action space $G_t\in[-5,5]\omega_m$ given the 
observation $O_t$ and is parameterized through $\mathcal{\pmb{\theta}}$. The 
$\mathbb{N}\times \mathbb{Z}$ independent trajectories, denoted as $\tau^{j}$ 
with the trajectory index $j=1,2,\ldots,\mathbb{N}\times \mathbb{Z}$, are 
collected with length $T=500$ (the number of time steps for each episode) and 
the step size $dt=0.01\omega_m^{-1}$. Each trajectory $\tau^{j}$ is a sequence 
of states (observations), actions, rewards, and next states (next observations): 
\begin{equation}
\tau^{j}=(O^{j}_0, G^{j}_0, R^{j}_0, O^{j}_1, \ldots, O^{j}_{T-1}, G^{j}_{T-1}, R^{j}_{T-1}), 
\end{equation}
which can be organized as a sub-trajectory tuple
\begin{equation}
   \tau^{j}_t = (O^{j}_t, G^{j}_t, R^{j}_t, O^{j}_{t+1}) 
\end{equation}
with the time stage index $t=0,1,\ldots,T-2$. At the terminal stage $t=T-1$, we 
have 
\begin{equation}
   \tau^{j}_{T-1}=(O^{j}_{T-1},G^{j}_{T-1},R^{j}_{T-1}). 
\end{equation}
For each sub-trajectory tuple $\tau^{j}_t$, the generalized advantage estimation 
(GAE)~\cite{schulman:2015} $\hat{A}^{j}_t$ uses a value function estimator:
\begin{equation}          
    \hat{A}^{j}_t = \delta^{j}_t + (\gamma\lambda)\delta^{j}_{t+1} + \ldots +(\gamma\lambda)^{T-t-1}\delta^{j}_{T-1}
\end{equation}
with
\begin{equation}
    \delta^{j}_t = R^{j}_t + \gamma V(O^{j}_{t+1};\pmb{\phi}) - V(O^{j}_t;\pmb{\phi}),
\end{equation}
where the value function $V(O^{j}_{t};\pmb{\phi})$ is utilized to score the 
quality of $O^{j}_t$ based on the accumulated reward and parameterized by 
$\pmb{\phi}$ and $\delta^{j}_t$ is the relative advantage of the current action 
selected by the stochastic policy $\pi(G^{j}_t|O^{j}_t;\pmb{\theta})$ with the 
discounted factor $\gamma\in(0,1)$ and hyperparameter $\lambda$ with typical 
value $\lambda=0.95$. Intuitively, $\hat{A}^{j}_t$ is utilized to numerically 
quantify the relative cumulative advantage of a certain action selected by the 
current stochastic policy from time $t$ to the terminal stage $T-1$, in which 
the future impact is included but regarded as less important than the 
corresponding previous one by the discount factor $\gamma\in(0,1)$. The 
finite-horizon discounted return $\hat{\mathcal{G}}^{j}_t$ is defined as
\begin{equation}\label{eq: return}
    \hat{\mathcal{G}}^{j}_t=\sum_{k=t}^{T-1}\gamma^{k-t}R^{j}_{k},
\end{equation}
which can be also obtained from the generalized advantage by 
\begin{equation}
    \hat{\mathcal{G}}^{j}_t = \hat{A}^{j}_t + V(O^{j}_t;\pmb{\phi}),
\end{equation}
where $\hat{\mathcal{G}}^{j}_t$ denotes the accumulated reward from time $t$ to 
the terminal stage in the discounted version.

The neural networks constituting the actor and critic are updated from 
minibatches with size $\mathbb{M}$ from $\mathbb{N}\times \mathbb{Z}\times T$ 
data points, consisting of the sub-trajectory $\tau^{j}_t$, the generalized 
advantage $\hat{A}^{j}_t$ and the return $\hat{\mathcal{G}}^{j}_t$ over $k=10$ 
epochs with the Adam algorithm. The typical batch size is 
$\mathbb{M}=\mbox{int}(\mathbb{N}\times \mathbb{Z}\times T/10)$. For each epoch, 
the critic parameters $\pmb{\phi}$ in the loss $L_{critic}(\pmb{\phi})$ and the 
actor parameters $\pmb{\theta}$ in the loss $L_{actor}(\pmb{\theta})$ need to be
updated to minimize the loss function over a random minibatch data. The mean 
square loss $L_{critic}(\pmb{\phi})$ about the target $\hat{\mathcal{G}}_i$ for 
the value function $V(O_i;\pmb{\phi})$ is 
\begin{equation}
    L_{critic}(\pmb{\phi})=\hat{\mathbb{E}}_{i}[(V(O_i;\pmb{\phi})-\hat{\mathcal{G}}_i)^2]
\end{equation}
and the clipped loss $L_{actor}(\pmb{\theta})$ is given by
\begin{align}
L_{actor}(\pmb{\theta})=\hat{\mathbb{E}}_i\left[-\textnormal{min}(r_i(\pmb{\theta})\hat{A}_i,\textnormal{clip}(r_i(\pmb{\theta}),[1-\epsilon,1+\epsilon])\hat{A}_i)\right],
\end{align}
where $\hat{\mathbb{E}}_{i}[\;\;]=\sum_{i=1}^{\mathbb{M}}[\;\;]_i/\mathbb{M}$ is 
the empirical average over a minibatch of the data and $[\;\;]_i$ denotes the 
$i$ th element of the minibatch with $i=0,1,\ldots,\mathbb{M}-1$, and the clip 
function $\textnormal{clip}(x,[\textnormal{min},\textnormal{max}])$ 
returns $x$ clipped to set limits: $\textnormal{min}\leq x\leq\textnormal{max}$. 
The probability ratio $r_i(\pmb{\theta})>0$ between the current and old policies 
is
\begin{align}
r_i(\pmb{\theta})=\frac{\pi_{\pmb{\theta}}(G_i|O_i)}{\pi_{\pmb{\theta}_{\textnormal{old}}}(G_i|O_i)}.
\end{align}
If the current policy is the same as the old policy, we have
$r_i(\pmb{\theta}_{\textnormal{old}})=1$. In general, the ratio 
$r_i(\pmb{\theta})$ needs to be away from the value one for the policy to be 
optimized. However, $r_i(\pmb{\theta})$'s deviating too much from the value one 
will result in many fast policy updates, possibly leading to instabilities
and even a collapse of the learning process. To avoid this, the clip function in 
the actor loss $L_{actor}(\pmb{\theta})$ can be utilized to remove the incentive 
for $r_i(\pmb{\theta})$ outside of the interval $[1-\epsilon,1+\epsilon]$ with 
typical clip range $\epsilon=0.2$, which decreases the updating speed of policy 
and improves the learning stability. 

Intuitively, the goal of RL is to maximize the cumulative reward. In the linear optomechanical system, the objective is to achieve the 
entangled Bell state as fast as possible or, as stipulated by the reward function, 
to achieve the optimal photon number $O_t\rightarrow 0.5$ and to maintain this for
as long as possible. When the RL agent converges to the optimal policy, the Bellman 
equation is satisfied~\cite{sutton:2018}, so the optimal value function satisfies
\begin{equation}
    V^*(O^{j}_t;\pmb{\phi}) = R^{j}_t + \gamma V^*(O^{j}_{t+1};\pmb{\phi}), 
\end{equation}
i.e., the optimal value function about $O^{j}_t$ is equal to the current reward 
plus the future discounted cumulative reward, in which $O^{j}_{t+1}$ is determined 
by the action selected by the optimal policy $\pi^{*}(G^{j}_t|O^{j}_t;\pmb{\theta})$. 
It guarantees that the agent makes the best possible decisions to maximize the 
rewards~\cite{sutton:2018}. Moreover, under the optimal policy, it means the zero 
generalized advantage $\hat{A}^{j}_t$, so the zero actor loss $L^{*}_{actor}$ is 
obtained. It is worth noting that the optimal value function is equal to the 
discounted accumulated reward from Eq.~\eqref{eq: return}:
\begin{equation}
    V^*(O_i;\pmb{\phi})=\hat{\mathcal{G}}_i,
\end{equation}
which also gives zero critic loss $L_{critic}(\pmb{\phi})$. In the online training 
process, the RL agent trained as described is called the PPO agent, whose policy 
is randomly initialized and will gradually converge to the optimal one under the 
described training scenarios to achieve the maximum accumulated reward. 
Physically, this enables the entangled Bell state to be created and stabilized.
For online testing, the optimized policy is no longer updated and only one 
quantum environment is involved.

\begin{figure*} [ht!]
\centering
\includegraphics[width=\linewidth]{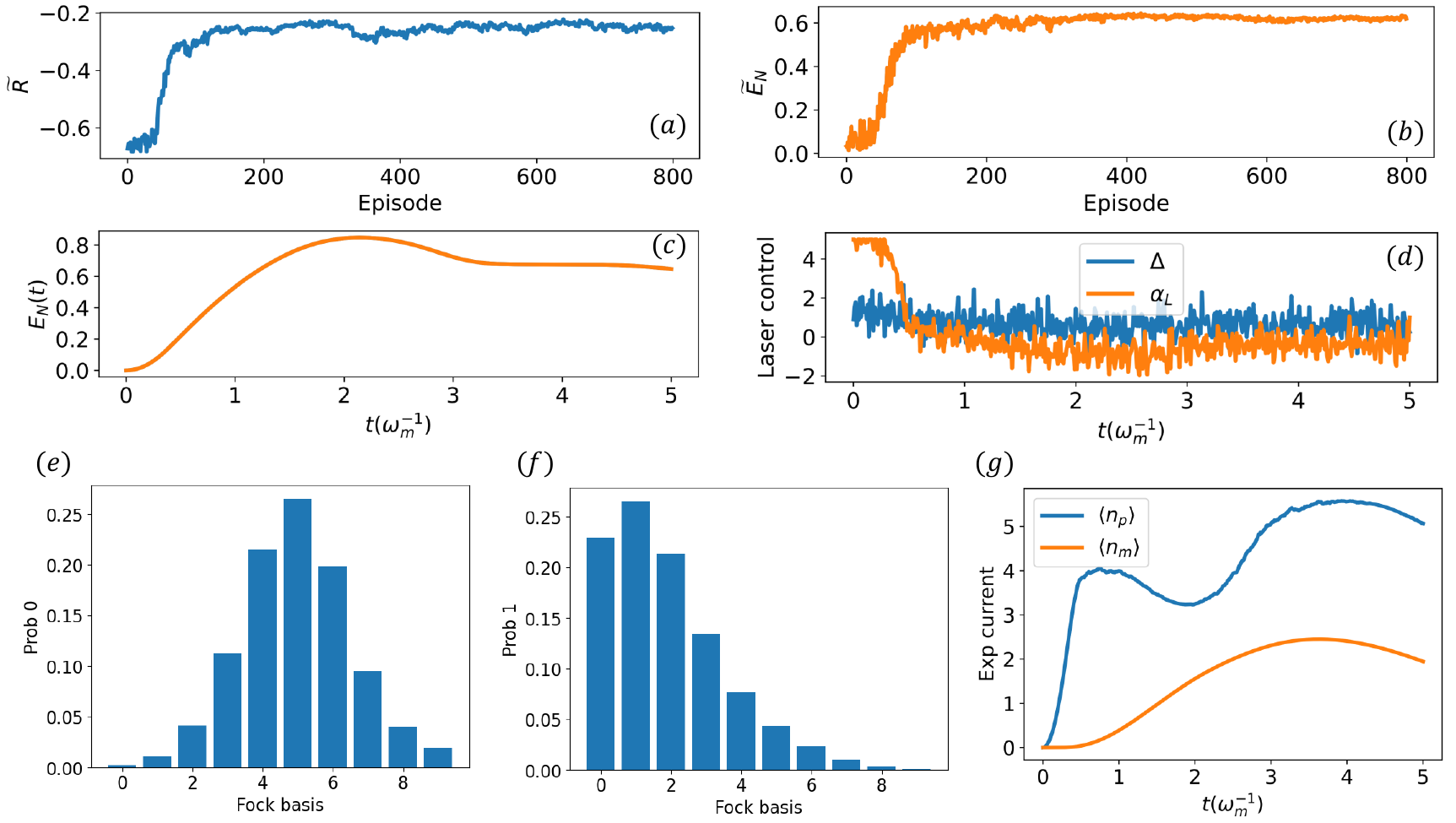}
\caption{A detailed account of the target-generating phase in RL control of open 
optomechanical systems with nonlinear photon-phonon interaction in the framework of 
Lindblad master equation. Nonlinear interaction of strength $g_0=0.2\omega_m$ creates 
the target entanglement $E_{N}\sim \log2$ optimized by the PPO agent from vacuum 
states with $|\psi\rangle = |00\rangle$ with $10\times 10$ Fock bases. The dissipation 
rates to the vacuum bath are $\kappa=0.1\,\omega_m$ and $\gamma = 0.01\kappa$. The 
time-dependent control signal is the detuning $\Delta$ and the amplitude of the driven 
laser $\alpha_L$ within the range $\Delta,\alpha_L\in[-5,5]\omega_m$. In the training 
phase, observation is set as $E_{N}(t)$. (a,b) Trained $\widetilde{R}$ and 
$\widetilde{E}_N$ converge to some constant values. (c,d) Time-dependent series 
$E_{N}(t)$, where the driven laser signals are shown at the end of the training 
phase. (e,f) The corresponding coherent- and thermal-shape states expanded in the 
Fock basis at the end of the time of the selected training episode in (c,d). (g) The 
time evolution of the corresponding expected measurement current, including the 
expected number $\langle n_p\rangle$ of photons as well as the expected phonon 
number $\langle n_m\rangle$ in the Fock basis, where the time series 
$\langle n_p\rangle$(t) serves as the target to construct reward function in the 
target-utilization phase shown in Fig.~\ref{fig: noWCM_nonlinear_obser_Pn}.}
\label{fig: noWCM_nonlinear_obser_EN}
\end{figure*}

\begin{figure*} [ht!]
\centering
\includegraphics[width=\linewidth]{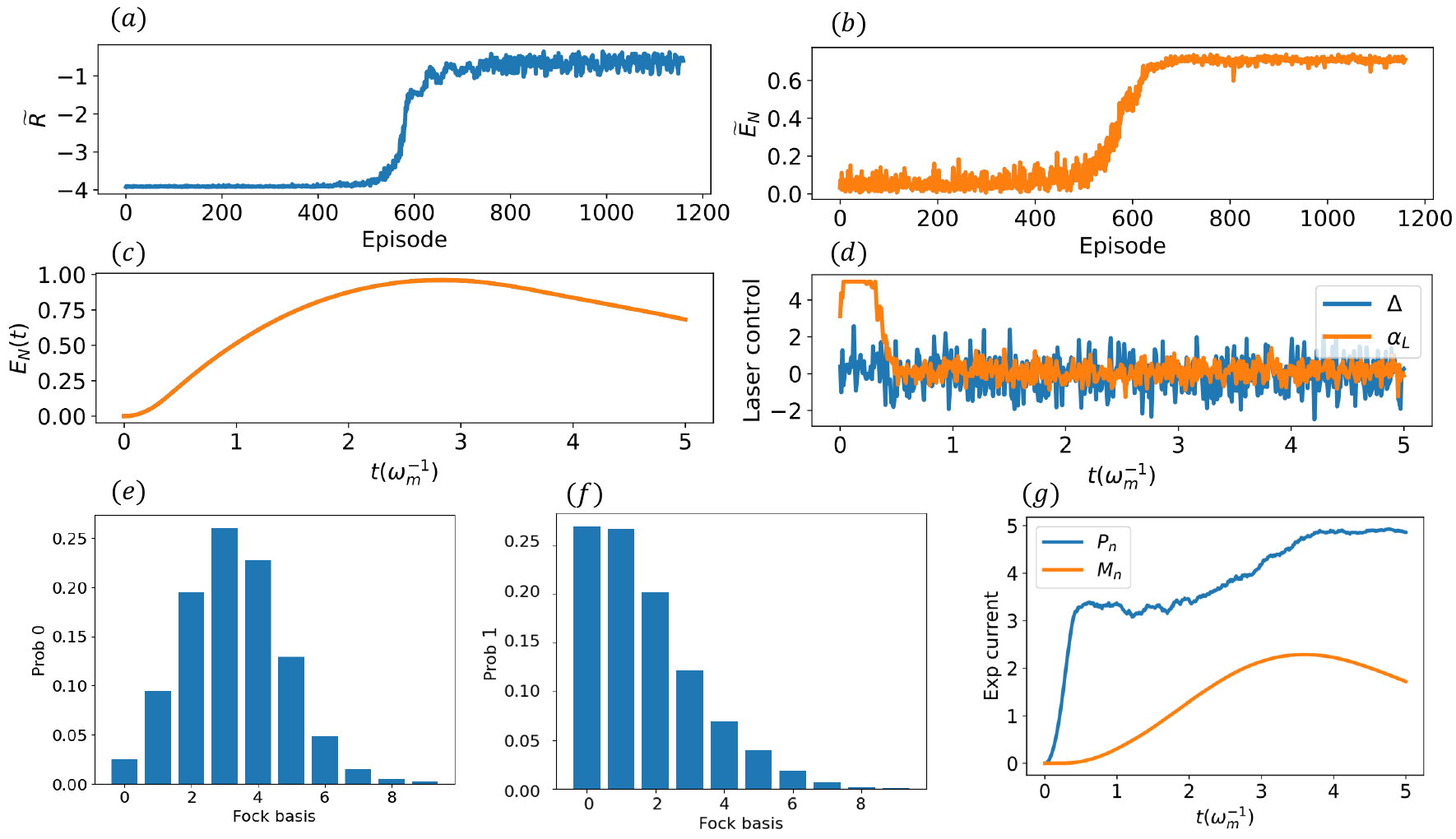}
\caption{A detailed account of the target-utilization phase in RL control of open 
optomechanical systems with nonlinear photon-phonon interaction in the framework 
of Lindblad master equation. The goal is to create the entanglement about 
$E_{N}\sim\log2$, for which the reward function is 
$R(t)=-|\langle n_p\rangle(t)-\langle n_p^{\textnormal{target}}\rangle(t)|$, 
determined by the target time series $\langle n_p^{\textnormal{target}}\rangle(t)$ 
from Fig.~\ref{fig: noWCM_nonlinear_obser_EN}(g). In this training configuration, 
the observation of the recurrent PPO agent is the expected photon number 
$\langle n_p\rangle(t)$. In spite of the observation being partial and incomplete, 
all results in (a-g) display similar behavior compared with the ones in 
Fig.~\ref{fig: noWCM_nonlinear_obser_EN}, where the entanglement quantity $E_{N}(t)$ 
is directly observed. However, such observation is currently not experimentally 
feasible. The setting and other parameters are the same as those in 
Fig.~\ref{fig: noWCM_nonlinear_obser_EN}.}
\label{fig: noWCM_nonlinear_obser_Pn}
\end{figure*}

\section{RL in nonlinear interactions by the Lindblad master equation}\label{sec:appendix_RL_nonlinear}

Figs.~\ref{fig: noWCM_nonlinear_obser_EN} and \ref{fig: noWCM_nonlinear_obser_Pn} 
display the case where the stochastic process in SME is removed so that the quantum 
dynamics are reduced to those governed by the Lindblad master equation, in which the 
decoherence part includes only the dissipation to the vacuum bath. In this setting, 
the nonlinear coupling represented by 
$\hbar g_0\hat{a}^{\dagger}\hat{a}(\hat{b}^{\dagger}+\hat{b})$ can still be exploited 
to create the entanglement. A caveat is that the process can simultaneously generate 
undesired high-level quantum states. A solution is to apply deep RL to create and 
stabilize the entanglement $E_{N}\sim\log2$, where the problem is how to control the 
excitation within a limited Fock basis. For this problem, a key is choosing the 
effective and experimentally feasible observation data.

Here, we describe in detail our two-step training process leading to a solution 
through the Lindblad master equation.  The first step is the target-generating phase,
in which numerical simulation is used to generate the observation and reward 
data, where the PPO agent observes the logarithmic negativity $E_{N}(t)$ directly 
and constructs the reward function combining the expected numbers of photons and 
phonons $R(t)=-|E_{N}(t)-\log2|-|\langle n_p\rangle(t)+\langle n_m\rangle(t)-a|/b$.
Figure~\ref{fig: noWCM_nonlinear_obser_EN} illustrates the target-generating phase, 
where the range of excited quantum states in the Fock basis is limited by the total 
number $\langle n_p\rangle+\langle n_m\rangle$ with the optimized hyperparameters 
$a=1$ and $b=40$. The target time series of the expected photon number is 
$\langle n_p^{\textnormal{target}}\rangle(t)$. The second step is the 
target-utilization phase, where the reward function is given by 
$R(t)=-|\langle n_p\rangle(t)-\langle n_p^{\textnormal{target}}\rangle(t)|$. The 
recurrent PPO will only observe the expected photon number $\langle n_p\rangle(t)$,
which is experimentally feasible. During the two-step training, the agent collects 
data from five parallel quantum optomechanical environments and updates the policy 
every five episodes.

More specifically, Figs.~\ref{fig: noWCM_nonlinear_obser_EN}(a) and 
\ref{fig: noWCM_nonlinear_obser_EN}(b) show that both the reward $\widetilde{R}$ and 
the logarithmic negativity $\widetilde{E}_N$ converge in time during the 
target-generating phase. The trained agent can create and stabilize the entanglement
as shown in Fig.~\ref{fig: noWCM_nonlinear_obser_EN}(c) controlled by the laser 
control signal shown in Fig.~\ref{fig: noWCM_nonlinear_obser_EN}(d). At the end of 
the time series, entanglement is produced from the coherent-(photon) and 
thermal-shape (phonon) Fock states as displayed in 
Figs.~\ref{fig: noWCM_nonlinear_obser_EN}(e) and 
\ref{fig: noWCM_nonlinear_obser_EN}(f). The corresponding target pattern 
$\langle n^{\textnormal{target}}_p\rangle(t)$ is demonstrated in 
Fig.~\ref{fig: noWCM_nonlinear_obser_EN}(g). In the target-utilization phase,
the target pattern $\langle n^{\textnormal{target}}_p\rangle(t)$ is time-dependent, 
which is difficult to learn if only MLPs are used. Here a single long short-term 
memory (LSTM) network is added after the MLPs in both the actor and critic network, 
so the whole neural-network architecture is able to handle the time-dependent data. 
Figure~\ref{fig: noWCM_nonlinear_obser_Pn} illustrates that, with only partial 
information extracted from the quantum optomechanical environment, the agent can 
steadily learn to create and stabilize entanglement. 

\section{Deep RL} \label{sec:appendix_DRL}

There are three main RL approaches~\cite{arulkumaran:2017} based, respectively, 
on (1) value functions, (2) policy search, and (3) a hybrid actor-critic method 
that employs both the value functions and policy search. Specifically, the 
actor-critic method uses the value function as a baseline for policy gradients,
based on a trade-off between variance reduction of policy gradients and bias 
associated with value functions. Incorporating deep neural networks as a 
powerful function approximator into RL to obtain the optimal value functions 
and the optimal policy leads to deep RL with the advantage of mitigating
the issues associated with high dimensionality (overcoming the curse of 
dimensionality). A difficulty with deep RL is the local minima in the 
neural-network dynamics with a large number of parameters when directly 
searching for the optimal policy~\cite{arulkumaran:2017}. A common solution is 
to use a trust region that prevents an updated policy from deviating too far 
from the previous policies, thereby guaranteeing monotonic enhancement in 
policy search. To implement this, the trust region proximal optimization (TRPO) 
method~\cite{wu:2017} can be exploited, which makes the advantage estimate in 
the surrogate objective function constrained by Kullback–Leibler (KL) 
divergence. The combination of TRPO and generalized advantage estimation (GAE) 
is one of the state-of-the-art RL techniques for continuous control.

\subsection{PPO agent} \label{subsec:appendix_PPO}

Proximal policy optimization (PPO)~\cite{schulman:2017} agent attains the data 
efficiency and reliable performance of TRPO with only first-order optimization 
through a novel objective with clipped probability ratios, which can be readily
implemented with reduced complexity. A typical online training process of PPO 
agent consists of the following steps:

{\em Step 1} - Initialization: initialize the actor $\pi(a|s;\pmb{\theta})$ and 
the critic $V(s;\pmb{\phi})$ with random parameters $\pmb{\theta}$ and $\pmb{\phi}$, 
respectively. Both the actor and critic are components of the PPO agent. The stochastic 
policy $\pi(a|s;\pmb{\theta})$ is the conditional probability distribution on action 
space $a$ given state $s$. The value function $V(s;\pmb{\phi})$ is utilized to score 
the quality of state $s$ based on the accumulated reward.

{\em Step 2} - Trajectory collection: The quantum state or quantum environment is 
initialized for the first episode or reset for the following episodes. The agent 
interacts independently with $\mathbb{N}$ parallel quantum optomechanical environments 
(identical and independent) using the current stochastic policy 
$\pi_{\pmb{\theta}}(a_t|s_t)$ at time $t$. After $\mathbb{Z}$ episodes, 
$\mathbb{N}\times \mathbb{Z}$ independent trajectories of length $T$ (the total time 
steps $T$ for each episode) are collected as sequences of states $s^{j}_{t}$, 
actions $a^{j}_{t}$, rewards $R^{j}_{t}$, and next states $s^{j}_{t+1}$, in which 
the sub-trajectory tuple $\tau^{j}_t$ is defined as 
\begin{align}
    \tau^{j}_t=(s^{j}_t, a^{j}_t, R^{j}_t,s^{j}_{t+1})
\end{align}
with the trajectory index $j=1,2,\ldots,\mathbb{N}\times \mathbb{Z}$ and the time 
index $t=0,1,\ldots,T-2$. At the terminal stage $t=T-1$, the following holds:
\begin{equation}
   \tau^{j}_{T-1}=(s^{j}_{T-1},a^{j}_{T-1},R^{j}_{T-1}).
\end{equation}
The sub-trajectory tuple $\tau^{j}_t$ can be utilized to calculate and evaluate 
the performance of the agent at each time stage $t$. The trajectory $\tau^{j}$ 
of length $T$ is the union of the sub-trajectory tuple $\tau^{j}_t$ in the form of 
\begin{align}
   \tau^{j}=\tau^{j}_0\cup \tau^{j}_1\cup ......\cup\tau^{j}_{T-1},
\end{align}
so the trajectory $\tau^{j}$ is given by 
\begin{align}
    \tau^{j}=(s^{j}_0, a^{j}_0, R^{j}_0, s^{j}_1, ... , s^{j}_{T-1}, a^{j}_{T-1}, R^{j}_{T-1}).
\end{align} 

{\em Step 3} - Generalized advantage estimator and return: Estimate the advantages 
for each sub-trajectory tuple $\tau^{j}_t$ in the collected trajectories. In 
particular, the generalized advantage estimation (GAE)~\cite{schulman:2015} uses 
a value function estimator:
\begin{align}
    \hat{A}^{j}_t = \delta^{j}_t + (\gamma\lambda)\delta^{j}_{t+1} + \ldots +(\gamma\lambda)^{T-t-1}\delta^{j}_{T-1},
\end{align}
with
\begin{align}
    \delta^{j}_t = R^{j}_t + \gamma V(s^{j}_{t+1};\pmb{\phi}) - V(s^{j}_t;\pmb{\phi}),
\end{align}
where $\delta_t$ is the relative advantage of the current action selected by the 
policy $\pi(a^{j}_t|s^{j}_t;\pmb{\theta})$ with the discounted factor $\gamma\in(0,1)$
and hyperparameter $\lambda$ (typical value $\lambda=0.95$). The generalized 
advantage $\hat{A}^{j}_t$ at time $t$ is the discounted cumulative advantage from 
time $t$ to the terminal stage $T-1$.

In episodic learning (policy update after each $Z$ number of episodes), the return 
$\hat{\mathcal{G}}(\tau^{j})$ is defined as the cumulative reward over the trajectory 
$\tau^{j}$, i.e., $\hat{\mathcal{G}}(\tau^{j})=\sum_{t=0}^{T-1}R^{j}_t$ with the 
time horizon $T$. For mathematical convenience, we use the discounted version, i.e., 
finite-horizon discounted return 
\begin{align} \nonumber
\hat{\mathcal{G}}(\tau^{j})=\sum_{t=0}^{T-1}\gamma^{t}R^{j}_{t}. 
\end{align}
It implies that future performance is also included but less important than the 
previous one. The return $\hat{\mathcal{G}}^{j}_t$ at each time step is the sum 
of the discounted reward from the current time $t$, 
\begin{align} \nonumber
\hat{\mathcal{G}}^{j}_t=\sum_{k=t}^{T-1}\gamma^{k-t}R^{j}_{k}, 
\end{align}
which can be also obtained from the generalized advantage:
\begin{align}
\hat{\mathcal{G}}^{j}_t = \hat{A}^{j}_t + V(s^{j}_t;\pmb{\phi}).
\end{align}

{\em Step 4} - Update of the actor and critic from minibatches of training data 
over $k$ epochs with Adam or stochastic gradient descent. For each epoch, we first
sample a random minibatch data set with size $\mathbb{M}$ from 
$\mathbb{N}\times \mathbb{Z}\times T$ data points, including the sub-trajectory 
tuple $\tau^{j}_t$, the corresponding advantage $\hat{A}^{j}_{t}$ and return 
value $\hat{\mathcal{G}}^{j}_t$. We then update the critic parameters $\pmb{\phi}$ 
by minimizing the loss $L_{critic}(\pmb{\phi})$ across all sampled minibatch data, 
which is given by
\begin{align}
L_{critic}(\pmb{\phi})=\hat{\mathbb{E}}_{i}[(V(s_i;\pmb{\phi})-\hat{\mathcal{G}}_i)^2],
\end{align}
where $\hat{\mathbb{E}}_{i}[\;\;]=\sum_{i=1}^{\mathbb{M}}[\;\;]_i/\mathbb{M}$ is 
the empirical average over a minibatch of data and $[\;\;]_i$ denotes the $i$th 
element of the minibatch with $i=0,1,...,\mathbb{M}-1$. After this, we update the 
actor parameters $\pmb{\theta}$ by minimizing the loss $L_{actor}(\pmb{\theta})$ 
given by
\begin{align}
L_{actor}(\pmb{\theta})=\hat{\mathbb{E}}_i\left[-\textnormal{min}(r_i(\pmb{\theta})\hat{A}_i,\textnormal{clip}(r_i(\pmb{\theta}),[1-\epsilon,1+\epsilon])\hat{A}_i)\right].
\end{align}
where the clip function $\textnormal{clip}(x,[\textnormal{min},\textnormal{max}])$ 
returns $x$ clipped to set limits, i.e., $\textnormal{min}\leq x\leq\textnormal{max}$. 
The probability ratio $r_i(\pmb{\theta})$ between the current and old policies is 
defined as
\begin{align}
r_i(\pmb{\theta})=\frac{\pi_{\pmb{\theta}}(a_i|s_i)}{\pi_{\pmb{\theta}_{\textnormal{old}}}(a_i|s_i)}.
\end{align}
If the current policy is the same as the old policy, we have 
$r_i(\pmb{\theta}_{\textnormal{old}})=1$. Otherwise, the ratio $r_i(\pmb{\theta})$ 
will be away from the value one to get the new optimized policy. The clip function 
in actor loss $L_{actor}(\pmb{\theta})$ is utilized to remove the incentive for 
$r_i(\pmb{\theta})$ outside of the interval $[1-\epsilon,1+\epsilon]$, which 
decreases the update speed of policy and improves the learning stability. 
   
{\em Step 5} - Repeating Steps (2-4) for a specified number of iterations 
or until convergence is achieved.

\subsection{Recurrent PPO agent}

In general, the dynamical process of RL is Markovian: the future depends only 
on the present state. While this suitably describes many processes, there are 
applications where a non-Markovian type of RL is required, e.g., partially 
observable Markov Decision Processes (POMDPs) or when the physical system to
be controlled is in a non-Markovian environment. Leveraging recurrent neural 
networks (RNNs) for memory-based agent learning provides a solution. In 
particular, a RNN can store past information as memory by introducing loops 
in the neural network, in contrast to, e.g., feed-forward neural networks where 
signals flow only from input to output in a one-way manner. However, 
conventional RNNs may not be able to efficiently connect the long past 
information to the present task, a problem known as gap sensitivity or 
vanishing gradient.

Long short-term memory (LSTM)~\cite{hochreiter:1997} is capable of learning 
long-term dependencies, thereby overcoming the vanishing gradient problem. The 
key component of LSTM is the cell state, which mimics a conveyor 
belt~\cite{Christopher:2015}. Information can be added or removed by the forget 
input, and output gates. Since the actor and critic networks underlying PPO are 
multilayer perceptrons (MLPs), e.g. a special class of the feed-forward neural 
networks with fully connected layers, applying LSTM after MLPs leads to a 
recurrent PPO agent, where MLPs are responsible for feature learning and LSTM 
contributes long-term history memorization. For a recurrent PPO, the state 
$s_{t}$ is replaced by observation $o_{t}$ and the hidden states $h_{t}$ with 
POMDPs~\cite{pleines:2022}.

\subsection{Details in deep RL}
Some details about the 
hyperparameter in the PPO agent are as follows: The discounted factor is 
$\gamma=0.99$, the parameter for the generalized advantage estimation(GAE) is 
$\lambda=0.95$, the clip range is set $\epsilon=0.2$, the maximum gradient is set 
to be $0.5$ and the learning rate is $0.5\times 10^{-3}$. Especially, GAE is 
normalized by subtracting its mean value and dividing by its standard deviation, 
the stochastic policy is based on the action noise exploration instead of the 
state-dependent exploration, and the value function is no clipping. Since the 
observation is the measurement current with large variance, it is necessary to 
apply a one-dimensional Gaussian filter from the Scipy package, of which the 
filter interval and standard deviation of the Gaussian kernel are listed in 
as bellow. In the process of variance reduction for WCM photocurrent, 
the measurement photocurrent is averaged over five trajectories (an independent ensemble) 
at each time step, and then averaged over the previously successive five time steps. 
Finally, the obtained data is filtered by the Gaussian kernel. In the updating phase, 
the network parameters from actor and critic are updated by Adam with the minibatch 
size, one-tenth of training data, and epochs $k=10$.

\begin{table} [ht!]
\caption{Gaussian filter with filter interval and standard deviation of the Gaussian kernel.}
\label{ta: Gaussian_filter}
\begin{tabularx}{\linewidth}{YYY}
\hline\hline
\specialrule{0em}{1pt}{1pt}
measurement rate & filter interval size & standard deviation\\
\specialrule{0em}{1pt}{1pt}
\hline
\specialrule{0em}{1pt}{1pt}
1.0 & 10 & 3.0 \\
\specialrule{0em}{1pt}{1pt}
0.7 & 10 & 4.5\\
\specialrule{0em}{1pt}{1pt}
0.5 & 10 & 6.0\\
\specialrule{0em}{1pt}{1pt}
0.3 & 20 & 6.0\\
\specialrule{0em}{1pt}{1pt}
0.1 & 100 & 24.0\\
\specialrule{0em}{1pt}{1pt}
0.05 & 150 & 48.0\\
\hline\hline
\end{tabularx}
\end{table}

The theoretical foundation of RL is rooted in Markov Decision Processes (MDPs). Consequently, any problem that can be described as an MDP can, in principle, be solved by RL. This implies that the next action taken depends only on the current state. However, to ensure robust performance, some RL algorithms utilize experience replay, where the policy is updated after accumulating a certain batch of experiences. Importantly, this does not mean RL controllers are non-Markovian or possess memory.

When combined with deep neural networks, RL becomes deep RL. Typically, deep RL employs deep feedforward neural networks to solve problems, adhering to the Markovian assumption. For our study, we use this network structure to handle linear interaction cases. For nonlinear cases, we observe that a recurrent PPO agent (MLPs + LSTM) outperforms a standard PPO agent (MLPs only). In the recurrent PPO setup, both the critic and actor networks are independent, consisting of MLPs followed by a single LSTM layer. Each network comprises fully connected layers of sizes $256\times 128\times 64$, followed by an LSTM with 256 hidden states. While there is no precise definition of how many time steps the LSTM can effectively "memorize" for a given task, we note that the 256 hidden states serve as a mechanism for capturing temporal dependencies. In the nonlinear case, the numerical density matrix has dimensions $10\times 10$, resulting in $100$ elements. This is slightly smaller than the number of hidden states in the LSTM, suggesting that the hidden states are sufficient for modeling the dynamics of the problem.
\bibliography{QOptoMech}

\end{document}